\documentclass[review,3p]{elsarticle}

\usepackage[colorlinks]{hyperref}
\usepackage{subfigure}
\usepackage{subfigmat}
\usepackage{color}
\usepackage{diagbox}

\usepackage{booktabs}
\usepackage{amsmath}
\usepackage{graphicx}

\usepackage{epstopdf}
\usepackage[capitalize]{cleveref}
\usepackage[ruled]{algorithm2e}
\usepackage[version=3]{mhchem}

\usepackage[english]{babel}
\usepackage{blindtext}
\usepackage{amsmath,amsfonts,stmaryrd,amssymb,theorem}

\makeatletter
\newcommand\rlarrows{\mathop{\operator@font \rightleftarrows}\nolimits}
\makeatother
\newcommand{\f}[2]{{\frac{#1}{#2}}}

\newcommand{\wh}[1]{{\widehat{#1}}}

\newcommand{\bfit}[1]{\textbf{\textit{#1}}}

 \def\uvec{{\mbox{\boldmath$u$}}}

 \def\0vec{{\mbox{\boldmath$0$}}}

 \def\phivec{{\mbox{\boldmath$\phi$}}}

 \def\Yvec{{\mbox{\boldmath$Y$}}}

 \def\Fmat{{\underline{\underline{{F}}}}}

\makeatletter
\long\def\MaketitleBox{%
  \resetTitleCounters
  \def\baselinestretch{1}%
  \begin{center}%
   \def\baselinestretch{1}%
    \Large\@title\par\vskip18pt
    \normalsize\elsauthors\par\vskip10pt
    \footnotesize\itshape\elsaddress\par\vskip36pt
    \end{center}%
}
\makeatother

\begin{document}

\title{Lyapunov exponent and Wasserstein metric as validation tools for assessing short-time dynamics and quantitative model evaluation of large-eddy simulation}

\author[Stanford]{Hao Wu}
\author[Stanford]{Peter C. Ma}
\author[ctr]{Yu Lv}
\author[Stanford]{Matthias Ihme}

\address[Stanford]{Department of Mechanical Engineering, Stanford University, Stanford, CA 94305, USA}
\address[ctr]{Department of Aerospace Engineering, Mississippi State University, Mississippi State, MS 39762, USA}

\maketitle

\section{Introduction}
The objective of this work is to address the need for assessing the quality of large-eddy simulations with particular application to turbulent reacting flows. Such assessments include the consideration of resolution requirements and the quantitative evaluation of the accuracy of simulation results. Over recent years, substantial efforts have been made towards the development of methods for assessing the quality of LES-calculations. These assessments require the considerations of the following aspects:
\begin{enumerate} 
 \item \bfit{Statistical analysis through comparisons of moments:} Common techniques for assessing the quality of LES-calculations are concerned with examining the statistical convergence of simulation results with respect to the mesh-refinement~\cite{POPE_NJP04,LESIQ,Vervisch2010778,Geurts2002}. These procedures follow the analysis of RANS-methods and the convergence evaluation of laminar flow calculations. In these methods, quality is defined as how well LES predicts quantities correctly with respect to a DNS solution, the latter of which is often not available. Current metrics used in determining LES quality address this problem by utilizing a statistical assessment of the solution, considering mean-flow quantities and higher-order statistical moments. Common statistical metrics include resolution requirements~\cite{POPE_NJP04}, index of resolution~\cite{LESIQ}, evaluation of the sub-grid scale velocity fluctuations~\cite{Vervisch2010778,Kemenov2011}, or the evaluation of the ratio of the turbulent viscosity to the molecular viscosity~\cite{Donghyun2007,Kemenov2011}. While these methods are useful to guide mesh-refinement, these exhibit three short comings. First, these methods rely on knowledge about subgrid-scale contributions, which are typically model and therefore not trivial to evaluate accurately. Second, these models provide information of the quality in a statistical sense, which requires the collection of statistical information, which is often associated with substantial computational cost. Third, LES is inherently unsteady that seeks to resolve turbulent dynamics, and these metrics only provide a direct assessment about the spatio-temporal scales that are resolved.
\item \bfit{Model validation:}
A main challenge in the development of models for turbulent reacting flows is the objective and quantitative evaluation of the agreement between experiments and simulations. This challenge arises from the complexity of the flow-field data, involving different thermo-chemical and hydrodynamic quantities that are typically provided from measurements of temperature, speciation, and velocity. This data is obtained from various diagnostics techniques, including nonintrusive methods such as laser spectroscopy and particle image velocimetry, or intrusive techniques such as exhaust-gas sampling or thermocouple measurements~\cite{HEITOR_MOREIRA_PECS1993,ECKBRETH_BOOK1996,KOHSE_HOINGHAUS_BARLOW_ALDEN_WOLFRUM_PCI2005,BARLOW_PCI2007}. Instead of directly measuring physical quantities, they are typically inferred from measured signals, introducing several correction factors and uncertainties~\cite{BARLOW_FRANK_KARPETIS_CHEN_CF2005}. Depending on the experimental technique, these measurements are generated from single-point data, line measurements, line-of-sight absorption, or multidimensional imaging at acquisition rates ranging from single-shot to high-repetition rate measurements to resolve turbulent dynamics~\cite{ALDEN_BOOD_LO_RICHTER_PCI2011}. This data is commonly processed in the form of statistical results from Favre and Reynolds averaging, conditional data, probability density functions, and scatter data. However, the diversity of data, it makes it difficult to fully utilize this data for a complete model assessment.

\item \bfit{Dynamic description:} LES is inherently unsteady and a dynamic measure is desirable to further characterize the LES quality in representing the dynamic content of a simulation. This is particularly relevant for flows that are inherently transient. The key observation is that turbulence is a deterministic chaotic phenomenon, which is characterized by an aperiodic long-term behavior exhibiting high sensitivity to the initial conditions. Different approaches exist to measure and characterize a chaotic solution~\cite{Ruelle1979,Farmer1983,Eckmann1985,Hilborn1994,Boffetta2002}. On one hand, geometric approaches estimate the fractal dimension of the chaotic attractor, which, in turn, gives an estimate of the active degrees of freedom of the chaotic dynamical system. An accurate measure is the Hausdorff dimension \cite{Farmer1983}, which is often approximated by box counting, based on phase-space partitioning and correlation dimension based on time series analysis~\cite{Hilborn1994}. On the other hand, dynamical approaches estimate the entropy content of the solution, namely the frequency with which a solution visits different regions of an attractor, for example by the Kolmogorov-Sinai entropy.
\end{enumerate}

The objective of this contribution is to introduce methods that enable the quantitative analysis of LES. The first metric introduces the \emph{Lyapunov exponent}~\cite{NASTAC_LABAHN_MAGRI_IHME_PRF2017} as a metric for the assessment of the dynamic content of LES-calculations. Specifically, the Lyapunov exponent provides a measure of chaos in turbulent flows, and represents the rate of separation, and its reciprocal is closely related to the predictability horizon of a chaotic solution. In particular, the Lyapunov exponent, $\lambda$, is amenable to a simple physical interpretation:  If a system is chaotic, given an infinitesimal initial perturbation to the solution, two trajectories of the system separate in time exponentially until nonlinear saturation. The average exponential separation is the Lyapunov exponent. A solution is typically regarded as being chaotic if there exists at least one positive Lyapunov exponent. The Lyapunov exponent is  (i) a robust indicator of chaos, (ii)  a global quantity describing the strange attractors, and (iii) relatively simple to calculate~\cite{Leith1972,Kida1990,NoninfinitesimalPerturbations,Boffetta2002,ConvexErrorGrowthRate}. In addition, there are several benefits of using this method over statistical methods. First, the  Lyapunov exponent can be used on transient simulations where a statistically stationary flow is not present and the ability to determine the resolved and unresolved turbulence fluctuations may not be possible. Secondly, calculating the  Lyapunov exponent can be accomplished quickly for arbitrary meshes or geometries and is independent of any closure models. Finally, using the analysis of the  Lyapunov exponent additional information on local turbulence and sensitivity to domain size and shape can be obtained. 

The second metric introduced in this work is the probabilistic \emph{Wasserstein metric}~\cite{JOHNSON_WU_IHME_CF2017} representing a measure for the quantitative evaluation of combustion models. Complementing currently employed comparison techniques, this metric possesses the following attractive properties. First, this metric directly utilizes the abundance of data from unsteady simulations and high-repetition rate measurement methods. Rather than considering low-order statistical moments, this metric is formulated in distribution space. As such, it is thereby directly applicable to scatter data that are obtained from transient simulations and high-speed measurements without the need for data reduction to low-order statistical moment information. Second, this metric is able to synthesize multidimensional data into a scalar-valued quantity, thereby aggregating model discrepancies for individual quantities. Third, the resulting metric utilizes a normalization, thereby enabling the objective comparison of different simulation approaches. Fourth, this method is directly applicable to sample data that are generated from scatter plots, instantaneous simulation results or reconstructed from statistical results. Fifth, the Wasserstein metric enables the consideration of conditional and multiscalar data, and is equipped with essential properties of metric spaces. Finally, this metric is a companion tool to previously established methods for validation of LES and instantaneous CFD-simulations.

The remainder of this paper is organized as follows: Section~\ref{SEC_ANALYSIS_TOOLS} provides a details discussion of the Lyapunov exponent (in Sec.~\ref{SSEC_LYAPUNOV_EXP}) and the Wasserstein metric (discussed in Sec.~\ref{SSEC_WASSERSTEIN_METRIC}). The experimental configuration, LES-model, and computational setup are presented in Sec.~\ref{SEC_CONFIG}, and results of both reacting and non-reacting simulations are examined in Section~\ref{SEC_RES}. The paper is concluded in Section~\ref{SEC_CONC}.
\section{\label{SEC_ANALYSIS_TOOLS}Analysis Tools}
\subsection{\label{SSEC_LYAPUNOV_EXP}Lyapunov Exponent for Assessing Dynamic Content}
This section introduces the Lyapunov exponent as a dynamic metric for assessing the short-time predictability of LES-calculations. Specifically, the Lyapunov exponent, $\lambda$, provides a measure of chaos in turbulent flows, and represents the rate of separation; its reciprocal is closely related to the predictability horizon of a chaotic solution. As such, this metric provides direct information about the dynamic content that is contain in LES.
\subsubsection{Dynamical system representation}\label{subsec:dyn_system}
A turbulent flow can be represented as a dynamical system,
    \begin{equation}
    \dot{\phivec}(t) = \Fmat(\phivec(t))\;,
    \end{equation}
    with initial conditions $\phivec(t=t_0) = \phivec_0$; $\Fmat$ is the set of bounded differentiable flow equations, $\dot{\phivec}$ denotes the temporal derivative of the state vector $\phivec$. For a general chemically reacting flow, $\phivec$ contains the velocity vector ($\uvec$), pressure ($p$), density ($\rho$), and vector of species mass fractions ($\Yvec$): $\phivec=(\uvec,p,\rho,\Yvec)^T$. The solution $\phivec(t)$ belongs to a vector space $H$, called the phase space. In the finite-dimensional case, $H=\mathbb{R}^N$, where $N\in\mathbb{N}$. In the infinite-dimensional case $H$ is a Hilbert space. The fluid dynamics problems studied are infinite-dimensional because they are governed by PDEs.  However, they are characterized by the existence of a bounded set, called strange attractor, because they are dissipative systems. This means that the turbulent solution lies in a fractal set with finite dimension~\cite{Temam1997}. Moreover, after numerical discretization, the phase space necessarily becomes finite-dimensional. Hence, the fluid systems are considered finite-dimensional in this paper.
    
\begin{figure}[!hbt!]
        \centering
        \includegraphics[width=0.6\textwidth]{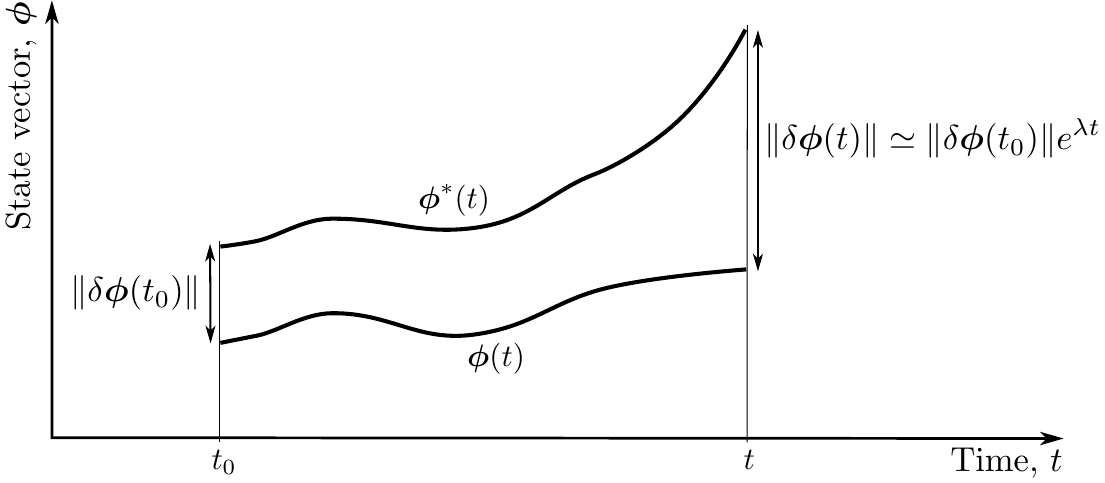}
        \caption{Separation of two slightly different solutions. The initial divergence is exponential and the growth rate is the Lyapunov exponent, $\lambda$.}
        \label{fig:trajectory}
    \end{figure}

Consider two initial conditions $\phivec_0$ and $\phivec^*_0$, which are infinitesimally distanced, $\phivec^*_0-\phivec_0=\delta\phivec_0$ (see Fig.~\ref{fig:trajectory} for a schematic illustration). The temporal evolution of the separation of the two trajectories, $\delta\boldsymbol{\phi}(t)$, in the tangent space, obeys the linearized dynamical equation $\delta\dot{\phi}_i(t) =\sum_{j=1}^{N_D}\frac{\partial F_i}{\partial\phi_j}\delta\phi_j(t_0)$, where $i=1,2,\ldots,N_D$, with $N_D$ being the number of degrees of freedom of the system, i.e., the dimension of the phase space. In the present study, $N_D\sim\mathcal{O}(10^7)$ for most cases. 
    
Under ergodicity, Oseledets~\cite{Oseledets1968} proved that there exists an orthonormal basis $\{{\bf{e}}_j\}$ in the tangent space such that the solution can be expressed by a modal expansion, $\delta\phivec(t)=\sum_{j=1}^{N_D}\alpha_j{\bf{e}}_je^{\lambda_jt}$, where the coefficients $\alpha_j$ depend on the initial condition $\delta\boldsymbol{\phi}(t_0)$. Mathematically, $\alpha_j=\langle{\bf e}_j,\delta\boldsymbol{\phi}(t_0)\rangle$, where the angular brackets denote an inner product. The exponents $\lambda_1\geq\lambda_2\geq...\geq\lambda_{N_D}$ are the Lyapunov exponents. Customarily, the maximal Lyapunov exponent, $\lambda_1$, is referred to as Lyapunov exponent and the subscript is omitted ($\lambda\equiv\lambda_1$). In the phase space, the modal expansion describes the deformation of a $N_D$-dimensional sphere of radius $\delta\phivec(t_0)$ centered at $\phivec(t_0)$ into an ellipsoid with semi-axes along the directions ${\bf{e}}_j$. Therefore, the Lyapunov exponents provide the stretching rates along these principal directions. Thus, given an infinitesimal initial perturbation to the solution, $\delta\boldsymbol{\phi}(t_0)$, the two trajectories of the system separate in time exponentially as~\cite{Boffetta2002} 
    \begin{equation} \label{Lyapunov_Growth_Rate}
    \lVert \delta\boldsymbol{\phi}(t) \rVert \simeq \lVert \delta\boldsymbol{\phi}(t_0) \rVert e^{\lambda t} , 
    \end{equation}
    where $\lVert\cdot\rVert$ is a norm. Figure \ref{fig:trajectory} illustrates the significance of the Lyapunov exponent. The predictability time, $t_p$, of the system for infinitesimal perturbations is then defined as the inverse of the Lyapunov exponent 
    \begin{equation} 
    \label{Lyapunov_Time}
    t_p  =  \ln{\left(\frac{\Vert\delta\boldsymbol{\phi}(t)\Vert}{\Vert\delta\boldsymbol{\phi}(t_0)\Vert}\right)}\frac{1}{\lambda}. 
    \end{equation}
\subsubsection{Calculation of the Lyapunov exponent as a separation growth rate}\label{subsec:algo}
The objective now is to utilize the Lyapunov exponent as an estimate for the rate of divergence of the Eulerian solution obtained by LES. From this information, a metric is proposed to measure how dynamically well-resolved the turbulent solution is. Using the Eulerian solution is a natural choice since most numerical simulations calculate Eulerian quantities. Growth rates of Eulerian fields have been used before in evaluating the error growth of weather models~ \cite{Leith1972,ConvexErrorGrowthRate}, finite perturbations of fully developed turbulence~\cite{NoninfinitesimalPerturbations} and decaying two-dimensional turbulence~\cite{Kida1990}. By observing that an Eulerian field can be regarded as a trajectory in an extended dynamical system~\cite{NoninfinitesimalPerturbations}, a practical method for obtaining the Lyapunov exponent is to perturb the initial field $\phivec(t_0)$ as 
    \begin{equation} 
    \label{Perturbation}
    \phivec^*(t_0) = \phivec(t_0) + \epsilon\left\|\boldsymbol{\phi}(t_0) \right\|\;, 
    \end{equation}
    where $\epsilon\ll1$, $\|\cdot\|\equiv\left(\frac{1}{V}\int_V \left(\cdot\right)^p dV\right)^{1/p}$ is the $L_p$-norm, and $V$ is the volume of the domain. The separation, also known as \emph{error} \cite{ConvexErrorGrowthRate}, is then measured by the $L_p$-norm of the subtracted Eulerian fields,
    \begin{equation}\label{eq:deltaphi}
    \left\lVert \delta \boldsymbol{\phi}\right\rVert=\left\lVert \boldsymbol{\phi}^* (t)-\boldsymbol{\phi}(t) \right\rVert.  
    \end{equation}
The separation behaves in accordance to Eq.~\eqref{Lyapunov_Growth_Rate}, thus, the Lyapunov exponent is computed as the linear slope of the natural logarithm of the separation versus time, $\lambda t=\ln{\left(\Vert\delta\boldsymbol{\phi}(t)\Vert/\Vert\delta\boldsymbol{\phi}(t_0)\Vert\right)}$. 
In the remainder of this work, the $L_2$-norm is chosen to measure the separation in Eq.~\eqref{eq:deltaphi}. 
    
If the process is ergodic, as assumed in this paper, the Lyapunov exponent is independent of the initial conditions as long as the nearly infinitesimal limit is satisfied \cite{Oseledets1968,Ruelle1979}.  
The procedure for calculating the Lyapunov exponent is given in Alg.~\ref{alg:lyapunovExp}. 

\begin{algorithm}[!hbt!]
1: Run numerical simulation until statistical convergence of the solution $\phivec$ is reached\\
2: Reset time to $t=t_0$ and save solution $\phivec$\\
3: At time $t=t_0$ compute $\phivec^*$ by perturbing the solution $\phivec$ following Eq.~\eqref{Perturbation}\\
4: Advance both solutions, $\phivec$ and $\phivec^*$, to some time $t$\\
5: Calculate the norm of the separation field $\delta\phivec$ from Eq.~\eqref{eq:deltaphi}\\
6: Take logarithm of separation norm, $\log_{10}(\left\Vert\delta\boldsymbol{\phi}\right\Vert)$ and calculate the Lyapunov exponent, $\lambda$, as slope of the linear region using linear regression\\
7: Normalize separation norm with saturation value to simplify inspection\\
\caption{Procedure for the evaluation of the Lyapunov exponent.}
	\label{alg:lyapunovExp}
\end{algorithm}

\subsubsection{Lyapunov metric for Dynamic-content Analysis of LES}
Compared to LES-quality metrics that rely on statistical information about turbulent kinetic energy or other flow-field quantities, the Lyapunov exponent intrinsically depends on the dynamic and chaotic nature of turbulence. While many turbulent flow systems in engineering are able to be time-averaged, some systems involve highly dynamic flows that cannot be averaged. For example, rare events are particularly difficult to simulate and capture, such as preignition, extinction or cycle-to-cycle variations in internal combustion engines. These rare events happen on a very small time scale, therefore, the simulations of these systems must be able to resolve the relevant dynamic scales to attempt to simulate these phenomena. As shown subsequently, the Lyapunov exponent saturates when the dynamical scales of the problem saturate. Specifically, as the spatio-temporal resolution approaches the smallest physical scales, the magnitude of the Lyapunov exponent reaches a plateau. This, in turn, provides a robust evaluation of the resolution requirements and spatial dimension of the computational domain in LES to capture the fundamental turbulent dynamics of rare deterministic events. 
    
One caveat in using the Lyapunov exponent is that its asymptotic value is not known \emph{a priori}; current results show that the Lyapunov time scale scales with the integral time scale for both the FHIT and jet cases. Iteration of resolution is likely required for more complex geometries and physics. The Lyapunov exponent is expected to be dependent on physical models and numerical discretization and can therefore be used as a sensitivity parameter and indicator to characterize their quality.
    
\subsection{\label{SSEC_WASSERSTEIN_METRIC}Wasserstein metric for Quantitative Analysis of Modeling Results}
This section the Wasserstein metric as a general methodology for quantifying discrepancies in simulations under consideration of multiple data-sources, including statistical moments, scatter data, and distributions.  Since multivariate distributions are commonly used to represent the thermo-chemical states in turbulent flows, this method is useful for the quantitative assessment of differences between a numerical simulation and experimental data of turbulent flames.

The dissimilarity between two points in the thermo-chemical state space can typically be quantified by its Euclidean distance. In the case of a 1D state space, e.g. temperature, the Euclidean distance is simply the absolute value of the difference. This metric serves the purpose well for the comparison among two deterministic measurements, which can indeed be fully represented by points in the state space. However, the Euclidean distance falls short when measurements are made on random data, e.g. states in a turbulent flow, in which case a single data point can no longer represent the measurement and shall be replaced by a distribution of possible outcomes. Consequently, the dissimilarity between two random variables shall be quantified by the ``distance'' between the corresponding probability distributions. 

Among many definitions of the ``distance" between distributions~\cite{gibbs2002choosing, Rubner2000}, the Wasserstein metric (also known as the Mallows metric) is of interest in this study. Several other methods for quantifying the difference between two distributions have been proposed. However, those methods are either limited to evaluating the equality of marginal distributions (e.g. Kolmogorov-Smirnov test) or are incomplete in that they fail to satisfy essential metric properties (e.g. Kullback-Leibler divergence). The interested reader is referred to the survey by Gibbs and Su~\cite{gibbs2002choosing} for a detailed comparison between the Wasserstein metric and other candidates. The Wasserstein metric represents a natural extension of the Euclidean distance, which can be recovered as the distribution reduces to a Dirac delta function. As such, this metric provides a measure of the difference between distributions that resembles the classical idea of ``distance". The resulting value can be judged and interpreted in the similar fashion as the arithmetic difference between two deterministic quantities.

This section is primarily concerned with the Wasserstein metric for discrete distributions in the Euclidean space. This allows us to focus on the concepts that are directly related to the practical calculation and estimation of this metric, and avoid the usage of measure theory without creating much ambiguity. The interested reader is also referred to books by Villani~\cite{villani2008optimal}, surveys by Urbas~\cite{urbas1998mass}, and the more recent lectures by McCann and Guillen~\cite{mccann2011five} for further details.
\subsubsection{Wasserstein metric for discrete distributions}
The Wasserstein metric is derived by considering two general discrete distributions, whose probability mass functions are:
\begin{equation}
\label{eq:disc_dist}
	f(x) = \sum_{i = 1} ^{n} f_i \delta(x-x_i) \;,\qquad	g(y) = \sum_{j = 1} ^{n^{\prime}} g_j \delta(y-y_i) \, ,
\end{equation}
where $\sum_{i=1}^{n} f_i = \sum_{j=1}^{n^{\prime}} g_i = 1$ and $\delta(\cdot)$ denotes the Kronecker delta function:
\begin{equation}
    \delta (x)= 
\begin{cases}
    1	& \text{if } x = 0 \, ,\\
    0	& \text{if } \text{otherwise} \, .
\end{cases}
\end{equation}
The unit cost of transport between $x_i$ and $y_j$ is defined to be the $p^\text{th}$ power of the Euclidean distance, i.e.~$c_{ij}= |x_i - y_j|^p$. In addition, the ``mass" of probabilities $f_i$ and $g_j$ is no longer restricted to be unity and the possible outcomes of the distributions, i.e.~$x_i$ and $y_j$, are points in an Euclidean space whose dimension is not restricted.

The $p^\text{th}$ Wasserstein metric between discrete distributions $f$ and $g$ is defined to be the $p^\text{th}$ root of the optimal transport cost for the corresponding Monge-Kantorovich problem:
\begin{equation}
\label{eq:wasserstein_metric}
\begin{aligned}
& W_p(f,g) & & = \underset{\Gamma}{\min} \bigg(
\sum_{i=1}^{n} \sum_{j=1}^{n^{\prime}} \gamma_{ij} |x_i - y_j|^p \bigg)^{1/p} \\
& \text{subject to}
& & \sum_{j = 1} ^{n^{\prime}} \gamma_{ij} = f_i \,,\quad \sum_{i = 1} ^{n} \gamma_{ij} = g_j \,,\quad\gamma_{i,j} \ge 0 \, .
\end{aligned}
\end{equation}
The constraints in Eq.~\ref{eq:wasserstein_metric} ensure that the total mass transported from $x_i$ and the total mass transported to $y_j$
match $f_i$ and $g_j$, respectively. 

\subsubsection{\label{SSEC_WASSERSTEIN_MC}Non-parametric estimation of $W_p$ through empirical distributions}
Two major difficulties arise in obtaining the Wasserstein metric of two multivariate distributions of thermo-chemical states. One is that the multivariate distributions are not readily available from either experiments or simulations. Instead, a series of samples drawn from these distributions is provided. The other is that there is no easy way of calculating $W_p$ for continuous multivariate distributions, especially for those of high dimensions. To overcome these problems, a non-parametric estimation of $W_p$ is devised using empirical distributions. 

An empirical distribution is a random discrete distribution formed by a sequence of independent samples drawn from a given distribution of interest. Let $(\mathcal{X}_1,\, \ldots,\, \mathcal{X}_n)$ be a set of $n$ independent random samples obtained from a continuous multivariate distribution $f$. The empirical (or fine-grained~\cite{POPE_BOOK00}) distribution $\wh{f}$ is defined to be
\begin{equation}
\label{eq:emp_dist}
	\wh{f}_n(x) = \frac{1}{n}  \sum_{i = 1} ^{n} \delta(x-\mathcal{X}_i) \, ,
\end{equation}
which is a discrete distribution with equal weights. The empirical distribution is random as it depends on random samples, $\mathcal{X}_i$. The samples may be obtained from experimental measurements or generated from a given distribution using, for instance, an acceptance-rejection method~\cite{RUBINSTEIN_KROESE_BOOK2008}. Calculation of $W_p$ between two empirical distributions is identical to the method discussed in Sec.~\ref{SSEC_WASSERSTEIN_METRIC}. Its procedure and cost is independent of the dimensionality of the distributions. 

The empirical distribution converges to the original distribution. Most importantly, in the context of this study, is the convergence in the Wasserstein metric~\cite{bickel1981some}. More specifically, the Wasserstein metric between empirical and original distributions converges to zero in probability. Given the metric property of $W_p$:
\begin{equation}
\label{eq:wp_metric_prop}
\left|W_p(\wh{f}_n, \wh{g}_{n^\prime}) - W_p(f, g)\right| \le W_p(\wh{f}_n, f) + W_p(\wh{g}_{n^\prime}, g) \, ,
\end{equation}
such convergence ensures that the Wasserstein metric between two empirical distributions also converges in probability to that of the actual distribution. In addition, the mean rates of convergence for empirical distributions have also been established~\cite{horowitz1994mean,fournier2015rate}, giving the upper bound on the expectation, $E\big(W_p^p(\wh{f}_n,\, f)\big),$ as $n$ increases. For the similar reason, the convergence rate for the non-parametric estimation follows. The exact rates depend on the dimensionality and regularity conditions of the distributions. For details on the convergence rate for more general cases, the interested reader is referred to the work by Fournier and Guillin~\cite{fournier2015rate}. 

In the case of $d$-dimensional distributions that have sufficiently many moments, we have:
\begin{equation}
	E\left(W_2^2(\wh{f}_n, f)\right) \le C \times
\begin{cases}
    n^{-1/2}		& \text{if } d < 4 \, ,\\
    n^{-1/2}	\log(1+n) & \text{if } d = 4 \, , \\
     n^{-2/d} & \text{if } d > 4 \, ,
\end{cases}
\end{equation}
where the value of $C$ depends on the distribution $f$ and is independent of $n$. By combining this result with Eq.~\ref{eq:wp_metric_prop}, we obtain the following convergence rate for the non-parametric estimation of $W_2$:
\begin{equation}
\label{e:rate_of_convergence_w2}
	E\left(W_2(\wh{f}_n, \wh{g}_{n^\prime}) - W_2(f, g)\right)^2 \le C \times
\begin{cases}
    n_{*}^{-1/2}		& \text{if } d < 4 \, ,\\
    n_{*}^{-1/2}	\log(1+n_{*}) & \text{if } d = 4 \, , \\
    n_{*}^{-2/d} & \text{if } d > 4 \, ,
\end{cases}
\end{equation}
where $n_{*} = \min(n,\, n^\prime)$. With this, all necessary results that ensure the soundness of using $W_p(\wh{f}_n, \wh{g}_{n^\prime})$ as an estimator for $W_p(f, g)$ are presented. 

In addition to the rate of convergence, statistical inference and further quantification of uncertainties for the non-parametric estimation of $W_p$ can also be performed. For instance, the magnitude of the uncertainty in Eq.~\ref{e:rate_of_convergence_w2} and the corresponding confidence interval can be estimated via the method of bootstrap~\cite{efron1979bootstrap,chernick2011bootstrap}. In particular, it was shown that the $m$-out-$n$ bootstrap~\cite{bickel1996m} can be applied to the Wasserstein metric~\cite{sommerfeld2016inference,dumbgen1993nondifferentiable}.
\subsubsection{\label{SSEC_SML}Statistically most likely reconstruction of distributions}
So far, the evaluation of the Wasserstein metric using sample data as empirical distribution function has been described. However, the usage of the Wasserstein metric is not limited to results reported in such fashion. In the following, the computation of the Wasserstein metric from statistical results will be discussed. This versatility is important to the metric, given the fact that the conventional practice of reporting only the statistics, predominantly first and second moments, is still the prevailing one. 

The procedure of applying the Wasserstein metric to statistical results is to first reconstruct the multivariate distribution from statistical results. An empirical distribution is then sampled to compute the Wasserstein metric following the method described in Sec.~\ref{SSEC_CALC_PROC}. The reconstruction of a continuous PDF from a set of known statistical models can be performed using the concept of the statistically most likely distribution (SMLD)~\cite{POPE_JNT79,IHME_PITSCH_PRT1_CF_2008}. The SMLD of a $d$-dimensional random variable is defined to be the distribution that maximizes the relative entropy, given a prior distribution $g(\mathbf{x})$, under a given set of constraints:
\begin{equation}
\label{eq:SMLD}
\begin{aligned}
& f_\text{SML}(\mathbf{x}) 
& & = \underset{f(\cdot)}{\operatorname{argmax}} \int_{\mathcal{R}^d} f(\mathbf{x}) \ln \left(\frac{f(\mathbf{x})}{g(\mathbf{x})}\right) d\mathbf{x} \\
& \text{subject to} 
& & \int_{\mathcal{R}^d} f(\mathbf{x}) d\mathbf{x} = 1\,, \\
& & & \int_{\mathcal{R}^d} \mathbf{T}(\mathbf{x}) f(\mathbf{x}) d\mathbf{x} = \overline{\mathbf{t}} \,. \\
\end{aligned}
\end{equation}
Here, $\mathbf{T}(\mathbf{x})$ is the set of statistical moments that are selected as constraints, and $\overline{\mathbf{t}}$ is the vector of corresponding values obtained from the data. The type of the so obtained distribution is dictated by the form of the constraints~\cite{lisman1972note}. For instance, if the multivariate distribution is constructed using only the first and second moments with a uniform prior, a multivariate normal distribution is obtained. In addition, if only the marginal second moments are given while the cross moments are not, the obtained multivariate normal distributions are uncorrelated. More specifically, if 
\begin{equation}
\begin{aligned}
g(\mathbf{x}) &\equiv 1 \, , \\
\mathbf{T}(\mathbf{x}) & = [\mathbf{x}_1,\mathbf{x}_2,\ldots,\mathbf{x}_d, \mathbf{x}_1^2, \mathbf{x}_2^2,\ldots, \mathbf{x}_d^2]\,,
\end{aligned}
\end{equation}
the SMLD becomes
\begin{equation}
f_\text{SML}(\mathbf{x}) = \prod_{k = 1}^{d} \frac{1}{\sqrt{2\pi \boldsymbol{\sigma}_k^2}} \exp{ \left\{ -\f{(\mathbf{x}_k - \boldsymbol{\mu}_k)^2}{ 2 \boldsymbol{\sigma}_k^2} \right\} } \, ,
\end{equation}
where $\boldsymbol{\mu}_k = \overline{\mathbf{x}}_k$ and $\boldsymbol{\sigma}_k^2 = \overline {\mathbf{x}^2_k} - \overline {\mathbf{x}_k}^2$.

After obtaining the SMLDs, the sets of samples can be drawn from them and the Wasserstein metric can be directly calculated. In addition, the metric property implies the following inequality
\begin{equation}
	|W_p(f_\text{SML},g_\text{SML}) - W_p(f,g)| \le W_p(f_\text{SML},f) +  W_p(g_\text{SML},g) \, .
\end{equation}
In other words, the difference in the Wasserstein metrics of the actual distributions, $f$ and $g$, and that of the reconstructed counterparts, $f_\text{SML}$ and $g_\text{SML}$, are bounded by the error of the SMLD reconstruction, which themselves can be quantified by the Wasserstein metric. In the current study, the set of constraints are limited to the marginal first and second moments, which are typically reported in the literature. More accurate reconstruction can be obtained by including high-order and cross moments. In addition, statistics of other forms may also be considered. For instance, $\beta$-distributions can be recovered for constraints in the form of $\overline{\ln(x)}$ and $\overline{\ln(1-x)}$, which is potentially more appropriate for conserved scalars such as mixture fraction.
\subsubsection{\label{SSEC_CALC_PROC}Calculation procedure}
\begin{figure}[!htb!]
\begin{center}
 \subfigure[\label{FIG_W2_DIST} PDF Sampling.]{\includegraphics[width = 0.45\textwidth]{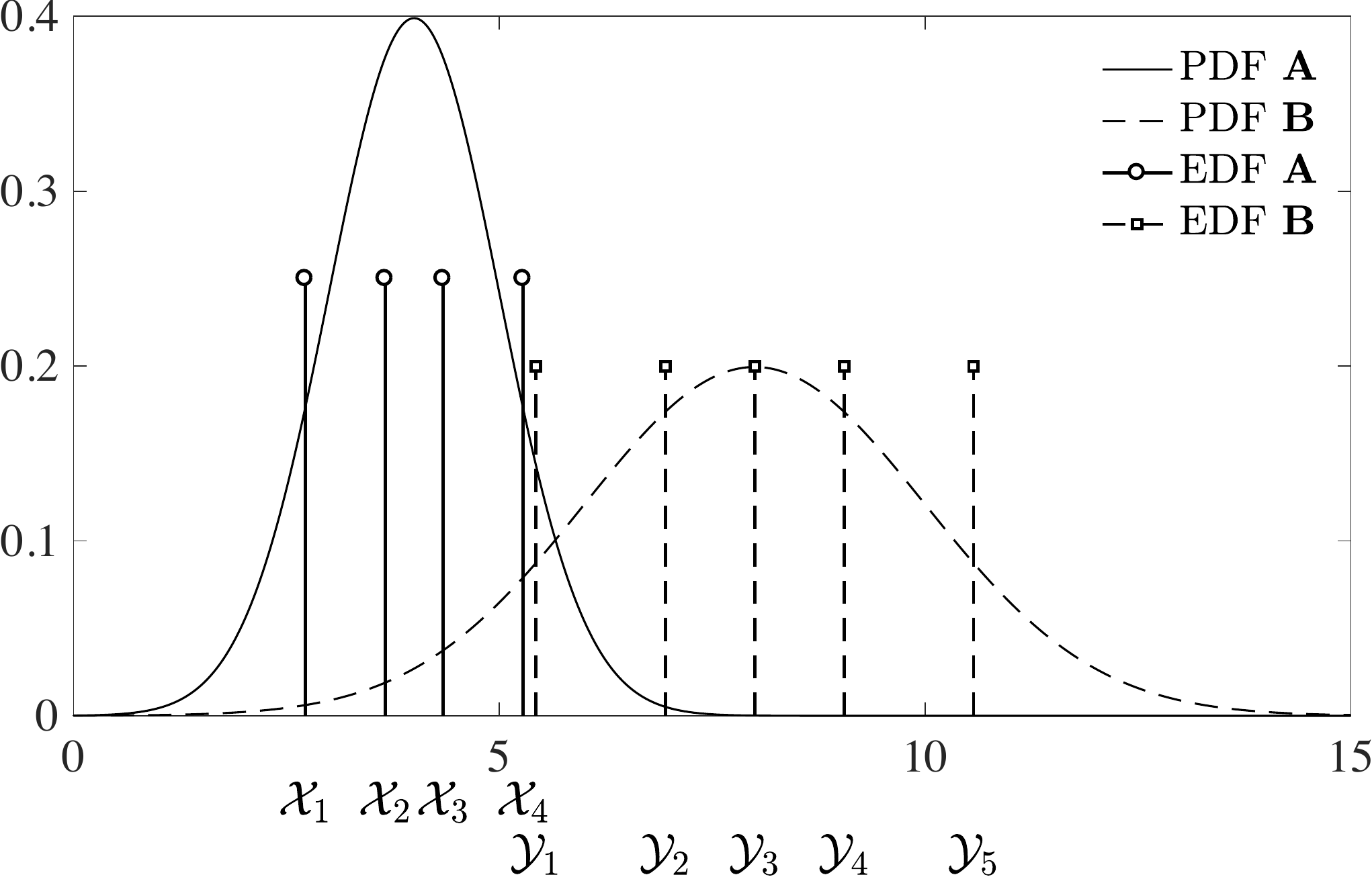}}\\
 \subfigure[\label{FIG_W2_FLOW}Transport matrix $\Gamma$.]{
    \begin{tabular}{|c||*{5}c|}
    \hline
    \diagbox{\small$\mathcal{X}$}{\small$\mathcal{Y}$} & 5.43 & 6.95 & 8.00 & 9.05 & 10.57\\\hline\hline
    2.72 & 0.15 & 0.10 &      &      & \\
    3.66 &      & 0.05 & 0.20 &      & \\
    4.34 &      & 0.05 &      & 0.20 & \\
    5.28 & 0.05 &      &      &      & 0.20\\\hline
    \multicolumn{6}{c}{}\\[-2ex]
    \end{tabular}
    \vspace*{4cm}
}
 \caption{\label{FIG_W2}Illustrative example of computing the Wasserstein matrix from two distribution functions: (a) The continuous PDFs are first sampled by empirical distribution functions (EDF, shown by discrete peaks, which are here down-sampled for illustrated purposes). The Wasserstein metric is then evaluated from the samples ${\mathcal{X}}_{i=1,\ldots,4}$ and ${\mathcal{Y}}_{i=1,\ldots,5}$ with (b) coefficient of transport matrix $\Gamma$, obtained from the solution to the minimization problem of Eq.~\ref{eq:wasserstein_metric}.}
\end{center}
\end{figure}

In the present study, the $2^{\text{nd}}$ Wasserstein metric is used. The computation of the metric can be realized by any general-purpose linear programming tool. For the present analysis, the program by Pele and Werman~\cite{Pele2009} is used. It calculates the Wasserstein metric as a flow-min-cost problem (a special case of linear programming) using the successive shortest path algorithm~\cite{ahuja1988network}. A pseudo code of the corresponding algorithm is given in Alg.~\ref{alg:Wasserstein_metric}, and Fig.~\ref{FIG_W2} provides an illustrative example for the evaluation of the Wasserstein metric. 

In this context, it is important to note that the input data to the Wasserstein metric are normalized to enable a direct comparison and enable a physical interpretation of the results. A natural choice is to normalize each sample-space variable by its respective standard deviation that is computed from the reference data set (for instance, the experimental measurements).
\begin{algorithm}[!hbt!]
	\SetKwInput{Input}{Input}
    \SetKwInput{Output}{Output}
	\SetKwInput{Preprocessing}{Preprocessing}
	\SetKwInput{Processing}{Processing}
	\Input{Two sets of $d$-dimensional data representing empirical distribution functions: $\mathcal{X}$, $\mathcal{Y}$, with lengths $n$ and $n^{\prime}$ from scatter data or sampled from continuous distribution functions}
    \Preprocessing{
	Normalize $\mathcal{X}$ and $\mathcal{Y}$ by standard deviation of one data set, $\boldsymbol{\sigma}_\mathbf{x}$
    }
	\For{i = $1: \, n$}{
		\For{j = $1: \,n^{\prime}$}{		
			Evaluate pair-wise distance matrix $c_{i,j} =\sum_{k=1}^{d} 
			\left(\mathcal{X}_{k,i}-\mathcal{Y}_{k,j}\right)^2$; }}
	Compute Wasserstein metric and transport matrix as solution to minimization problem of Eq.~\ref{eq:wasserstein_metric} using the shortest path algorithm by Pele \& Werman~\cite{Pele2009} with input $c_{i,j}$\\
	\Output{Wasserstein metric: $W_2$}
 	\caption{Pseudo code for evaluating the Wasserstein metric.}
	\label{alg:Wasserstein_metric}
\end{algorithm}

Suppose we have two sets of data with sample sizes of $n$ and $n^{\prime}$, respectively. Each sample represents a point in the thermo-chemical (sub)-space, e.g.\ $\mathbf{x} = [ Z,T,Y_{\ce{H2O}},\ldots ]$. The empirical distribution can be constructed  from each data set following Eq. \ref{eq:emp_dist}, where $f_i = 1/{n}$ and $g_i = 1/{n^{\prime}}$. The Wasserstein metric is then computed following the definition in Eq.~\ref{eq:wasserstein_metric}. The worst time complexity of the algorithm is $\mathcal{O} \left((n+n^{\prime})^3 \log(n + n^{\prime})\right)$. Note that the dimension of the thermo-chemical (sub)-space affects only the pair-wise distance between data points but not the definition or calculation of the metric.
\subsubsection{Remarks on the interpretation and usage of the Wasserstein metric $W_p$}
The Wasserstein metric is a natural extension of the Euclidean distance to statistical distributions. It enables the comparison between two multi-dimensional distributions via a single metric while taking all information presented by the distributions into consideration. The definition of $W_p$ in Eq.~\ref{eq:wasserstein_metric} can be viewed as the weighted average of the pair-wise distances between samples of two distributions. In the case of one-dimensional distributions, the obtained value of the metric shares the same unit as the sample data. For instance, if two distributions of temperature are considered, the corresponding Wasserstein metric in units of Kelvin can be interpreted as the average difference between the values of temperature from the two distributions. In the case of multi-dimensional distributions, each dimension is normalized before pair-wise distances are calculated. The choice of the normalization method is application-specific. In the present study, the marginal standard deviation is chosen. The so obtained $W_p$ represents the averaged difference, which is proportional to the marginal standard deviations, between samples from the two distributions. As such, a value of $W_p = 0.5$ can be interpreted as a difference between simulation and experimental data at the level of $0.5$ standard deviation. Although not considered in this study, additional turbulent-relevant information, such as space-time correlations, can be factored in via the extension of the phase space for the PDFs as performed by Muskulus and Verduyn-Lunel~\cite{muskulus2011wasserstein}.

Comparable samples need to be drawn from numerical simulations and experimental data to ensure a consistent comparison between the two distributions. This can be achieved by matching the sampling locations and frequencies between the two sources of data. Furthermore, experimental uncertainties can also be considered by either convolving the simulation data with the error distribution or through Bayes deconvolution of the experimental data~\cite{doi:10.1093/biomet/asv068,laird1978nonparametric}.

\section{\label{SEC_CONFIG}Experiment, Numerical Methods, and Computational Setup}
\subsection{Experimental Configuration: Volvo test rig}
\begin{figure}[!t!]
    \centering
    \includegraphics[width=0.65\textwidth]{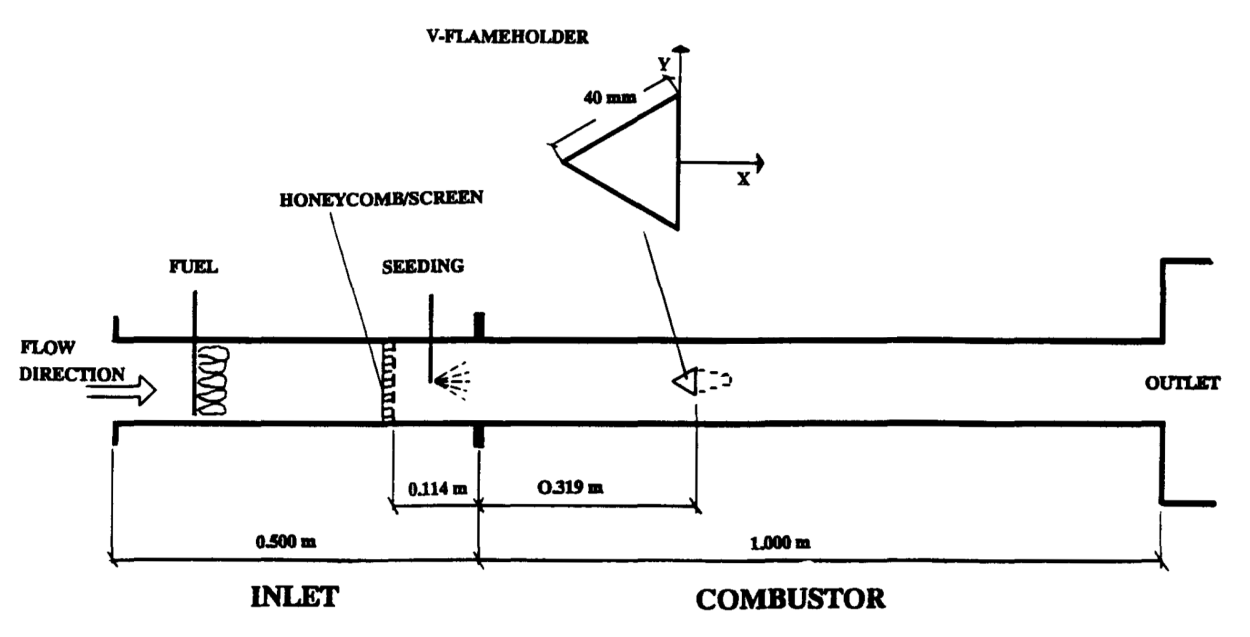}
    \caption{Schematic of the Volvo test rig (adapted from Sjunnesson~et~al.~\cite{ab1992alaa}). \label{fig:testrig}}
\end{figure}

A schematic of the Volvo test rig is shown in \cref{fig:testrig}. The configuration consists of a 1.5~m long rectangular duct which is 0.24~m wide by 0.12~m high with a choked air inlet. The apparatus is divided into the inlet and combustor sections, which are 0.5~m and 1.0~m in length, respectively. The inlet section is responsible for fueling, seeding, and flow conditioning. A multi-orifice critical flow injector is utilized for the injection of gaseous propane upstream of the inlet section for the reacting cases. Honeycombs located in the inlet section are used to control the turbulence level and mixture homogeneity. A 0.04~m equilateral triangle is mounted 0.682~m upstream from the exit spanning the width of the duct, and acts as the bluff-body flameholder. The flow exits the rectangular duct through a circular outlet.

Experimental measurements available for the Volvo test rig includes velocity data from Laser Doppler Anemometry (LDA)~\cite{sjunnesson1991lda}, and temperature and species data from Coherent Anti-Stokes Raman Scattering (CARS) technique~\cite{ab1992alaa}. Mean and root mean square (RMS) velocity data are available for axial profiles along the center-line and transverse profiles across the height of the combustor section at several axial locations, downstream the flameholder. 

The operating conditions under consideration are summarized in \cref{tab:conditions}. An air mass flow rate of 0.6~kg/s is considered. The Reynolds number of 48,000 is defined using the characteristic length of the bluff body ($D$~=~0.04~m) and bulk inlet velocity and viscosity at the temperature of 288~K. For the reacting case the equivalence ratio is 0.65, which gives an adiabatic flame temperature of 1784~K.

\begin{table}[!b!]
    \centering
    \begin{tabular}{c c c c c c}
        \toprule 
        Case & Re & $U_\text{bulk}$ [m/s] & $\phi$ & $T_\text{in}$ [K] & $T_\text{ad}$ [K]\\
        \midrule
        Non-reacting & 48,000 & 16.6 & 0.0 & 288 & - \\
        Reacting & 48,000 & 17.3 & 0.65 & 288 & 1784\\
        \bottomrule
    \end{tabular}
    \caption{Operating conditions. \label{tab:conditions}}
\end{table}

\subsection{Mathematical Model} \label{SEC_NUM}

The fully compressible Navier-Stokes equations are considered as the governing equations in this study. The Favre-averaged conservation equations of mass, momentum, total energy, and species, are written as follows:
\begin{subequations}
    \label{eqn:governingEqn}
    \begin{align}
        \frac{\partial \bar{\rho}}{\partial t} + \frac{\partial \bar{\rho} \widetilde{u}_j}{\partial x_j} &= 0\,,\\
        \frac{\partial \bar{\rho} \widetilde{u}_i}{\partial t} + \frac{\bar{\rho} \widetilde{u}_i \widetilde{u}_j}{\partial x_j} &= -\frac{\partial \bar{p}}{\partial x_i} + \frac{\partial}{\partial x_j} \left[ (\widetilde{\mu} + \mu_t) \left( \frac{\partial \widetilde{u}_i}{\partial x_j} + \frac{\partial \widetilde{u}_j}{\partial x_i} - \frac{2}{3} \delta_{ij} \frac{\partial \widetilde{u}_k}{\partial x_k} \right) \right] \,,\label{eqn:momentumequation} \\
        \frac{\partial  \bar{\rho} \widetilde{E}}{\partial t} + \frac{\bar{\rho} \widetilde{u}_j \widetilde{E}}{\partial x_j} &=
        \begin{aligned}[t]
            &\frac{\partial}{\partial x_j} \left[ \left(\widetilde{\frac{\lambda}{c_p}} + \frac{\mu_t}{\text{Pr}_t} \right) \frac{\partial \widetilde{h}}{\partial x_j} - \widetilde{u}_j \bar{p} + \widetilde{u}_i (\bar{\tau}_{ij} + \bar{\tau}^R_{ij}) \right] \\
            &+ \frac{\partial}{\partial x_j} \left[ \sum_{k = 1}^N \left(\bar{\rho}\widetilde{D}_k - \widetilde{\frac{\lambda}{c_p}} \right) \widetilde{h}_k \frac{\partial \widetilde{Y}_k}{\partial x_j} \right] \,,
        \end{aligned} \label{eqn:energyequation} \\
        \frac{\partial \bar{\rho} \widetilde{Y}_k}{\partial t} + \frac{\bar{\rho} \widetilde{u}_j \widetilde{Y}_k}{\partial x_j} &= \frac{\partial}{\partial x_j} \left[ \left( \bar{\rho}\widetilde{D}_k + \frac{\mu_t}{\text{Sc}_t} \frac{\partial \widetilde{Y}_k}{\partial x_j} \right) \right] + \bar{\dot{\omega}}_k\,,
    \end{align}
\end{subequations}
where $u_i$ is the i-th component of the velocity vector, $E$ is the total energy including the chemical energy, $C$ is the progress variable, $\mu$ and $\mu_t$ are the laminar and turbulent viscosity, $\lambda$ is the thermal conductivity, $D_k$ is the diffusion coefficient for the species $k$, $\dot{\omega}_k$ is the source term for species $k$, $\tau_{ij}$ and $\tau_{ij}^R$ are the viscous and subgrid-scale stresses, Pr$_t$ is the turbulent Prandtl number, and Sc$_t$ is the turbulent Schmidt number. An appropriate subgrid-scale model is needed for the computation of the turbulent viscosity. The system is closed by the ideal-gas law as the equation of state. The sub-grid terms associated with the equation of state are neglected in this study.

The unstructured LES-solver CharLES$^{\,x}$ is used in this study. This code has been extensively used for turbulent flow calculations \cite{hickey2013large, larsson2015incipient, wu2017mvp, ma2017framework, ma2017flamelet, lv2017underresolved}. A brief summary of the numerical schemes is provided here, and for more details the interested readers are referred to the work by Ham~et~al.~\cite{ham2004energy} and Khalighi~et~al.~\cite{khalighi2011unstructured}.

\begin{figure}[!b!]
    \centering
    \subfigure[4 mm]{
        \includegraphics[width=0.9\textwidth]{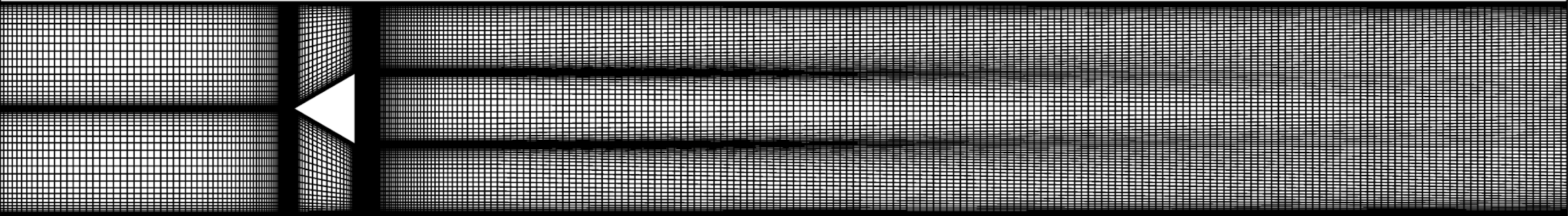}
    }
    \subfigure[2 mm]{
        \includegraphics[width=0.9\textwidth]{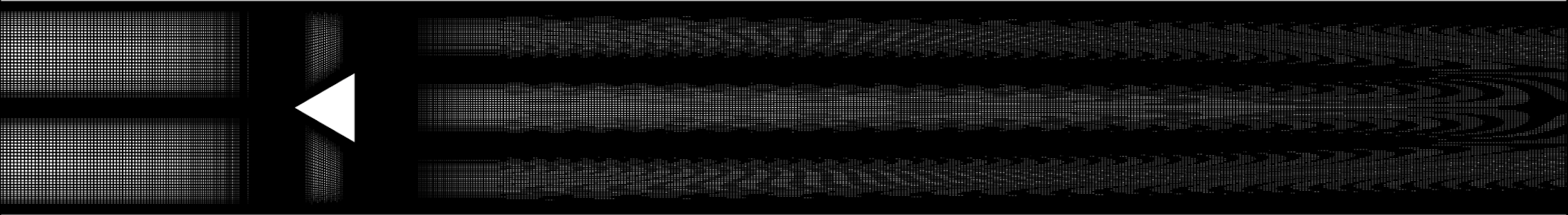}            
    }
    \subfigure[4 mm (top half flameholder)]{
        \includegraphics[trim={0 30.5cm 0 0},width=0.42\textwidth,clip]{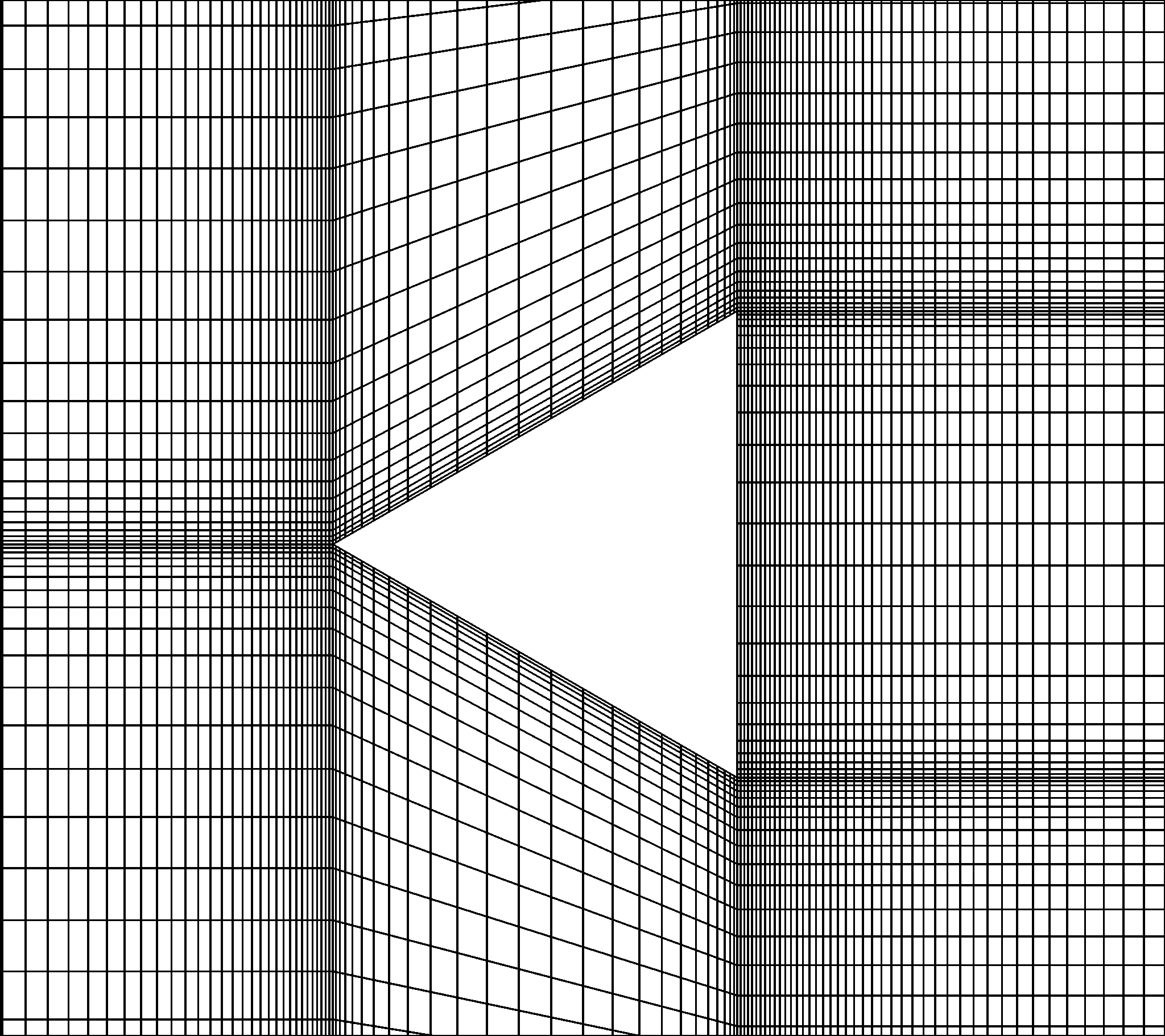}
    } \quad
    \subfigure[2 mm (top half flameholder)]{
        \includegraphics[trim={0 30.5cm 0 0},width=0.42\textwidth,clip]{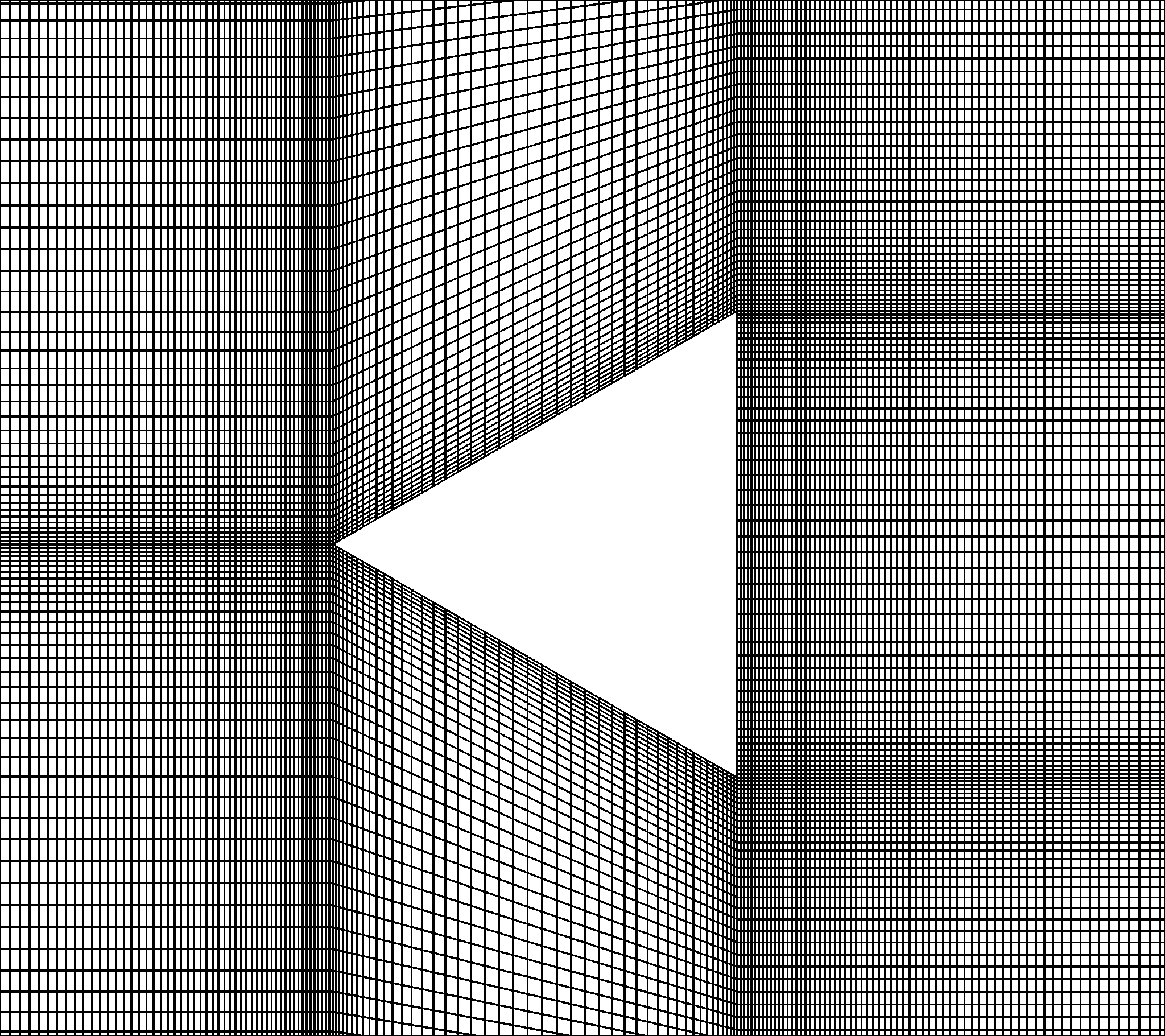}
    }
    \caption{Computational grids for the FV solver. \label{fig:meshfv}}
\end{figure}

A control-volume based FV approach is utilized for the discretization of the system of \cref{eqn:governingEqn}:
\begin{equation}
    \frac{\partial U}{\partial t} V_{cv} + \sum_f F^e A_f = \sum_f F^v A_f + S\Delta_{cv}\,,
\end{equation}
where $U$ is the vector of conserved variables, $F^e$ is the face-normal Euler flux vector, $F^v$ is the face-normal viscous flux vector, which corresponds to the r.h.s of \cref{eqn:governingEqn}, $S$ is the source term vector, $\Delta_{cv}$ is the volume of the control volume, and $A_f$ is the face area. The combination of a strong stability preserving 3rd-order Runge-Kutta (SSP-RK3) scheme \cite{gottlieb2001strong} and the semi-implicit ROK4E schemes is used for time advancement~\cite{WU_IHME_ROK4E_2016,wu2017application,WU_MA_IHME_2018}. CharLES$^{\,x}$ was developed based on a reconstruction-based FV scheme. Polynomials with maximum third-order accuracy are used to reconstruct the left- and right-biased face centroid values of the flow variables, and a blending between a central flux and a Riemann flux is computed based on the local grid quality. This flux computation procedure yields formally second-order accuracy and has maximum fourth-order accuracy on a perfectly uniform Cartesian mesh, without numerical dissipation. For computations of shock-related problems and cases where large gradients of flow variables are present, CharLES$^{\,x}$ utilizes a sensor-based hybrid Central-ENO scheme to minimize the numerical dissipation while stabilizing the simulation. For regions where shocks and large gradients are present, a second-order ENO reconstruction is used at the left- and right-biased face values, followed by an HLLC Riemann flux computation. A number of different hybrid sensors/switches are available. For this study, a \`relative solution" (RS) sensor is used which is based on the solution and reconstruction values of density~\cite{khalighi2011unstructured, ma2014supercritical}. An entropy-stable flux correction technique~\cite{ma2016entropy} is used to ensure the physical realizability of the numerical solution and to dampen non-linear instabilities.

Following Pope's criterion~\cite{pope2004ten}, Cocks~et~al.~\cite{cocks2013reacting} showed that a resolution of 3~mm is required so that 80\% of the turbulent kinetic energy can be fully resolved by the LES, although the criterion may not be sufficient for reacting cases. Based on this argument, three grids with increasing resolutions of 4~mm, 2~mm, and 1~mm are generated. The 4~mm and 2~mm grids are shown in \cref{fig:meshfv} along with zoomed-in regions near the flameholder. The grids are clustered near the wall, with a minimum wall spacing of 0.3~mm, and the mesh is nearly isotropic in the rest of the domain. A uniform grid is employed in the spanwise direction. The three grids amount to a total of 0.54, 4.17, and 24.6 million cells, respectively. 

A Vreman SGS model~\cite{Vreman2004} is used for the sug-grid terms. The sub-grid scale eddy viscosity is calculated as
\begin{equation}
    \mu_t = \bar{\rho} C \Delta_{cv}^{2/3} \left( \frac{\beta_{11}\beta_{22}-\beta_{12}^2+\beta_{11}\beta_{33}-\beta_{13}^2+\beta_{22}\beta_{33}-\beta_{23}^2}{\alpha_{ij}\alpha_{ij}} \right)^{1/2}\,,
\end{equation}
where $\alpha_{ij} = \partial \widetilde{u}_j / \partial x_i$, $\beta_{ij} = \alpha_{mi}\alpha_{mj}$, and $\Delta_{cv}$ is the local cell volume. The model constant $C$ is set to a value of 0.07 in this study.

The combustion process in the FV simulations is described by a two-step kinetic scheme as mentioned in the workshop document~\cite{mvpws2016}, which is adapted from Ghani et~al.~\cite{ghani2015longitudinal}. The laminar flame speed and the adiabatic flame temperature obtained by this mechanism are listed in Table~\ref{tab:flamespeed} in comparison with those from the GRI 3.0 mechanism~\cite{smith1999gri}. The difference in the flame speed is $11 \%$ and the flame temperature is virtually identical. 

\begin{table}[!b!]
    \centering
    \begin{tabular}{c c c c c c}
        \toprule 
        Mechanism & $S_\text{l}$ [m/s] & $T_\text{b}$ [K]\\
        \midrule
        2-step & 0.194 &  1805.8 \\
        GRI30 & 0.218 & 1806.6 \\
        \bottomrule
    \end{tabular}
    \caption{Laminar flame speed and the adiabatic flame temperature obtained by the two-step mechanism for the FV simulations in comparison to GRI30~\cite{smith1999gri}. \label{tab:flamespeed}}
\end{table}

The thickness of the flame is estimated to be 0.6 mm, which cannot be resolved by the meshes used in this study. Therefore, a dynamic thickened flame model~\cite{colin2000thickened,legier2000dynamically} is used to describe the flame turbulence interactions with the model of Charlette et al.~\cite{charlette2002power} for the sub-grid efficiency. 

\subsection{Computational setup}
\begin{figure}[!tb!]
    \centering
    \includegraphics[width=0.9\textwidth]{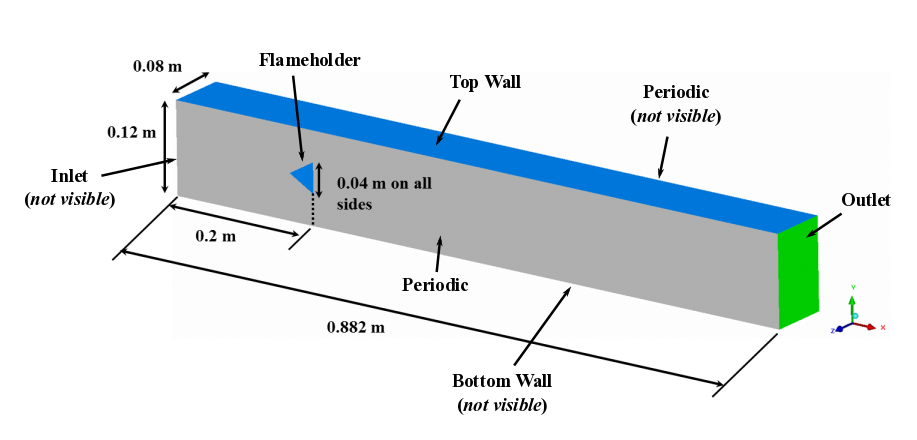}
    \caption{Computational domain and boundary conditions~\cite{mvpws2016}. \label{fig:domain}}
\end{figure}

The computational domain considered in this study is shown in \cref{fig:domain}, with a depth twice the width of the bluff body, corresponding to one third of the channel in the experiment. The full combustor section downstream of the flameholder is considered. The bluff body is placed 0.2~m from the inlet. Periodic boundary conditions are applied in the spanwise direction. A fixed mass flow rate of 0.2083 kg/s is employed at the inlet through characteristic boundary conditions applied by velocity, temperature and species. No turbulence profile is included for the inlet velocity and a plug-flow profile is adopted. Pressure of 1~bar is specified through characteristic boundary conditions at the outlet. No-slip adiabatic wall boundary conditions are employed at the flameholder and top and bottom walls, following the recommendation of the workshop.

\section{Results and  Discussions} \label{SEC_RES}
In this section, the reacting and non-reacting simulations are conducted, and results are presented in comparison with the experimental data. Subsequently, the analysis tools presented in Sec.~\ref{SEC_ANALYSIS_TOOLS} are applied and its versatility is examined.

\subsection{Statistical Flow-field Analysis}
\subsubsection{Non-reacting simulations}
\begin{figure}
    \centering
    \includegraphics[trim={2cm 26cm 2cm 18.5cm},width=0.95\textwidth,clip]{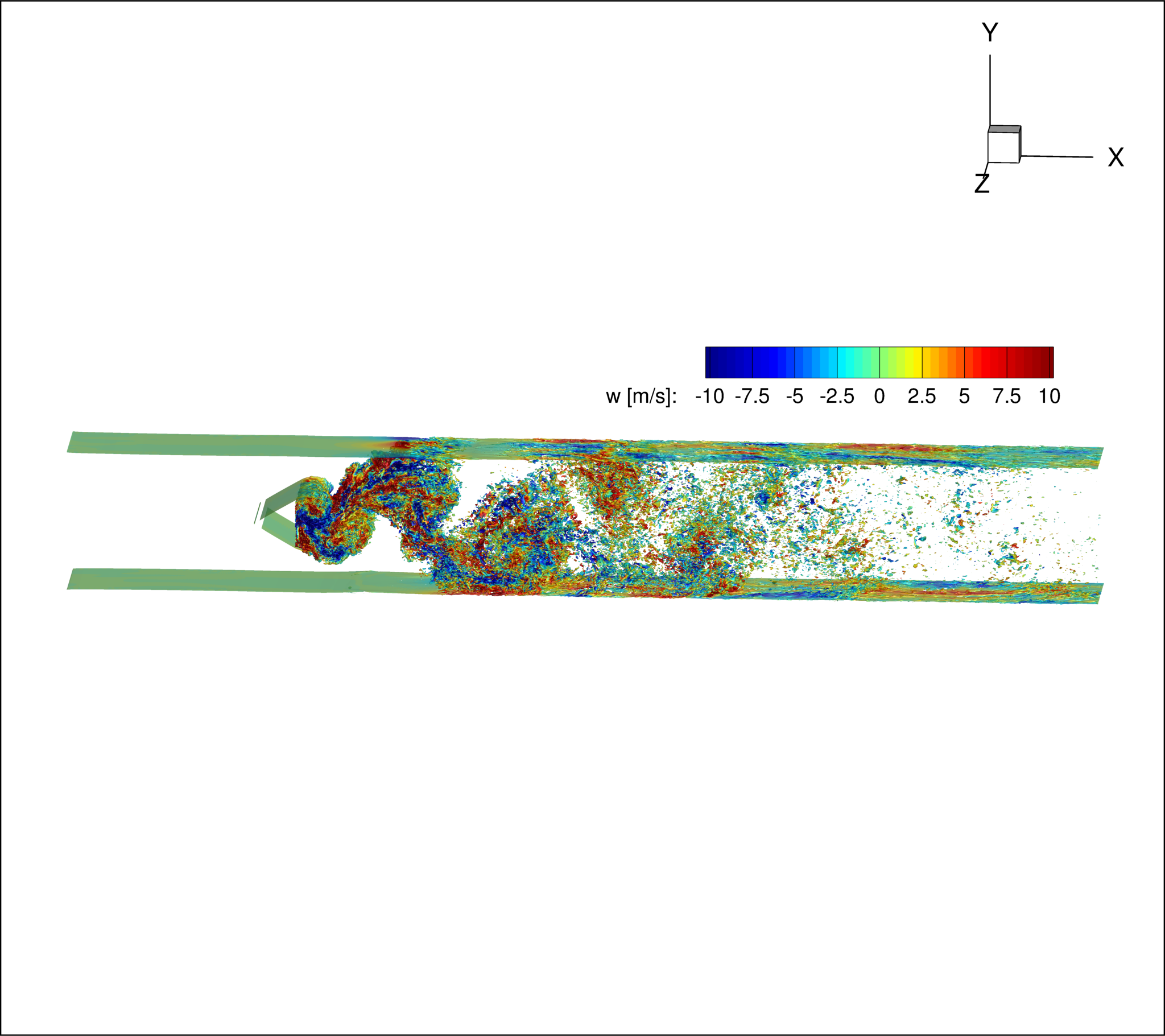}
    \caption{Iso-surfaces of vorticity magnitude of 4000~s$^{-1}$ colored by the spanwise velocity on the 1~mm grid for the non-reacting case from the FV solver. \label{fig:vorticity}}
\end{figure}

The non-reacting case is simulated with the FV solver using three different resolutions, namely 4~mm, 2~mm, and 1~mm. For the sake of validating the sub-grid scale models utilized and establishing grid convergence study. \Cref{fig:vorticity} shows the vortical structures exhibited downstream of the bluff body from the simulation using the 1~mm grid. Iso-surfaces of vorticity magnitude of 4000~s$^{-1}$ colored by the spanwise velocity are displayed. It can be seen that the von-Karman type shedding is generated behind the flameholder due to the instabilities produced by the flow separations. The large vortices then break up into smaller vortices and the flow becomes more turbulent further downstream. It is also important to observe the interactions between the vortices and the wall. It has been pointed out by Cocks~et~al.~\cite{cocks2013reacting} that the wall vortices play a major role inestablishing three-dimensional vortex breakdown further downstream of the domain.

\begin{figure}
    \centering
    \includegraphics[trim={30 0 40 0},width=0.48\textwidth,clip]{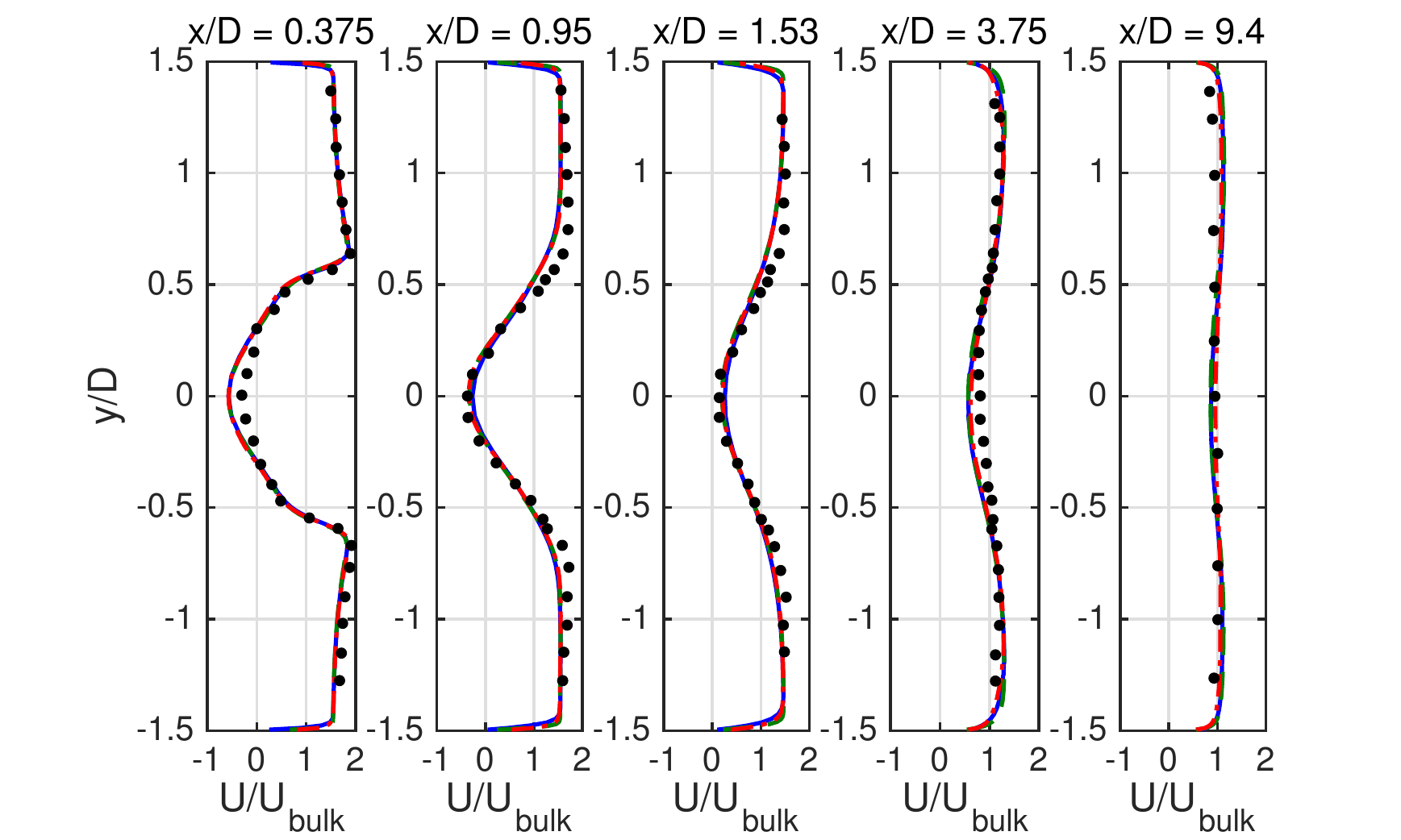}
    \includegraphics[trim={30 0 40 0},width=0.48\textwidth,clip]{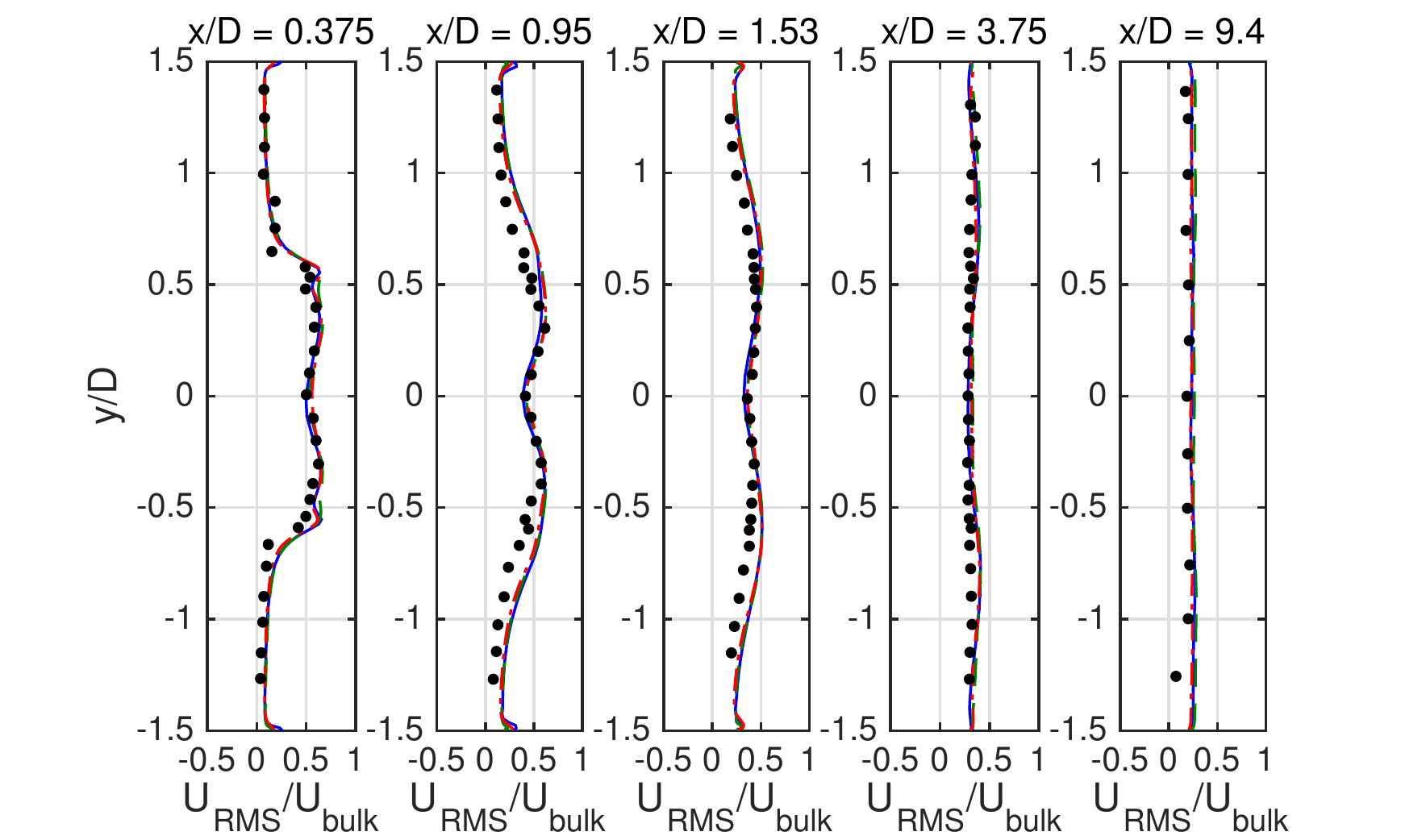}
    \caption{Normalized mean (left) and RMS (right) axial velocity profiles on three grids (solid line~\textendash\,~4~mm, dashed line~\textendash\,~2~mm, dash-dotted line~\textendash\,~1~mm) in comparison with measurements (dots), at several axial locations for the non-reacting case from the FV solver. \label{fig:velocityu}}
\end{figure}

\begin{figure}
    \centering
    \includegraphics[trim={30 0 40 0},width=0.48\textwidth,clip]{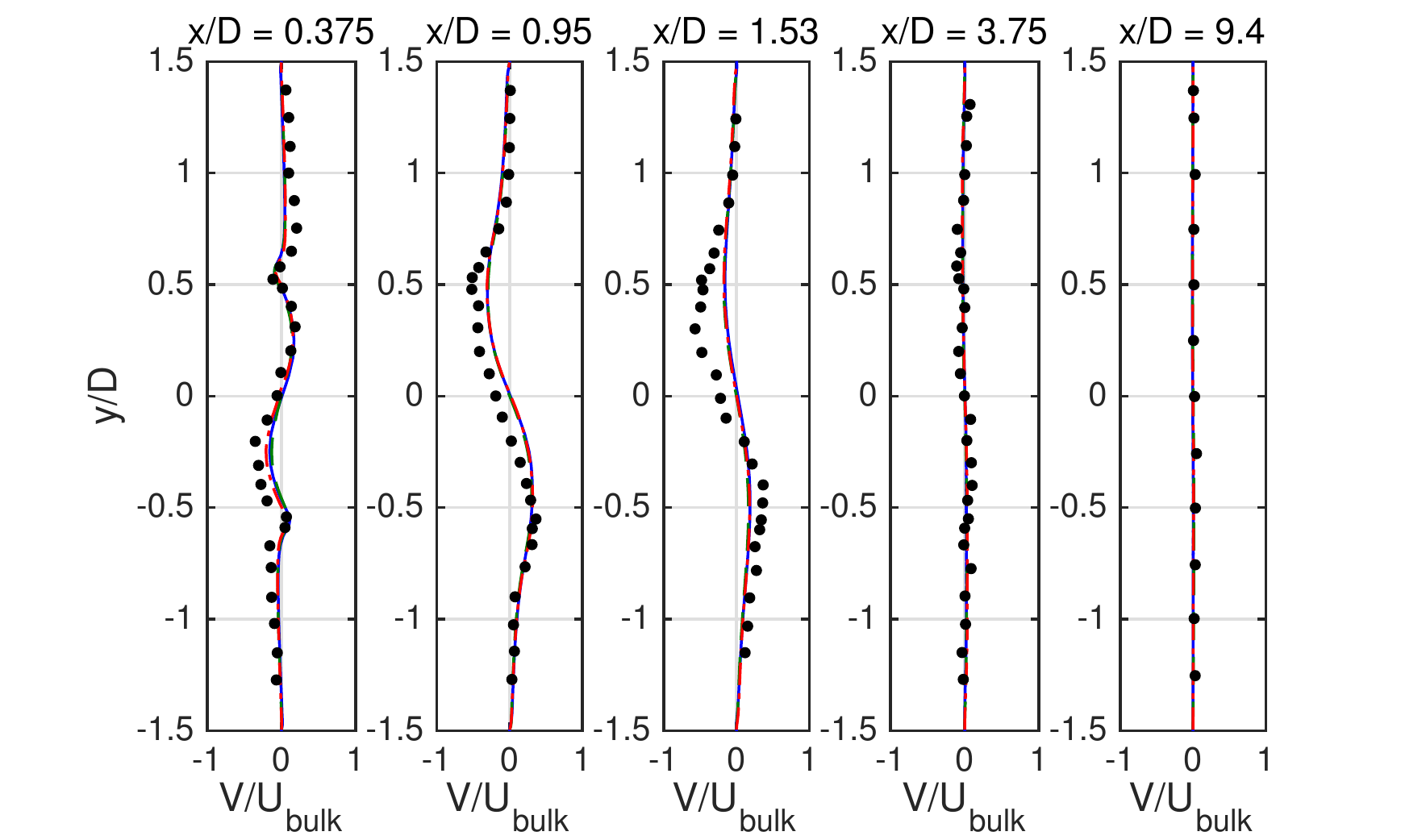}
    \includegraphics[trim={30 0 40 0},width=0.48\textwidth,clip]{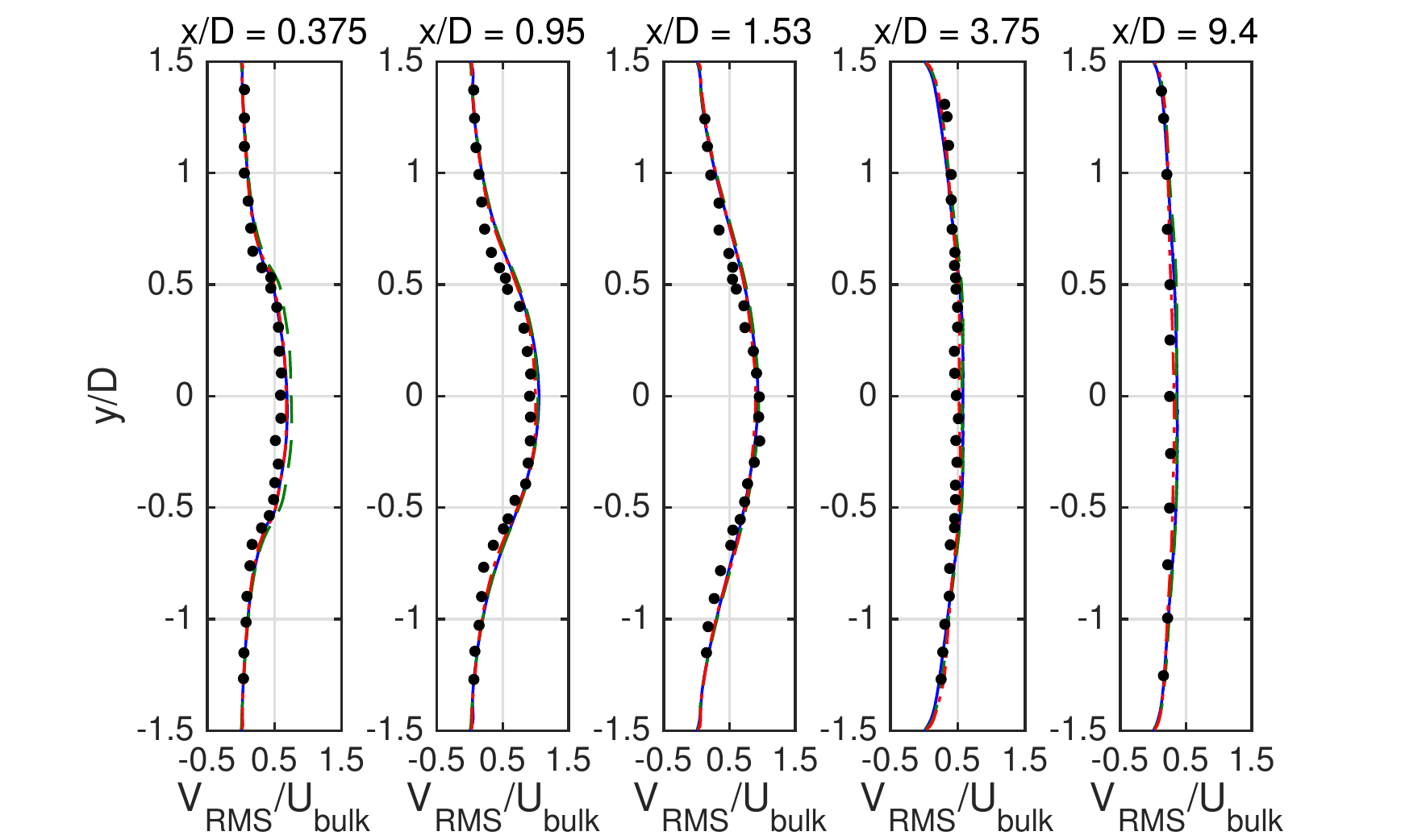}
    \caption{Normalized mean (left) and RMS (right) transverse velocity profiles on three grids (solid line~\textendash\,~4~mm, dashed line~\textendash\,~2~mm, dash-dotted line~\textendash\,~1~mm) in comparison with measurements (dots), at several axial locations for the non-reacting case from the FV solver. \label{fig:velocityv}}
\end{figure}

\begin{figure}
    \centering
    \includegraphics[trim={30 0 40 0},width=0.48\textwidth,clip]{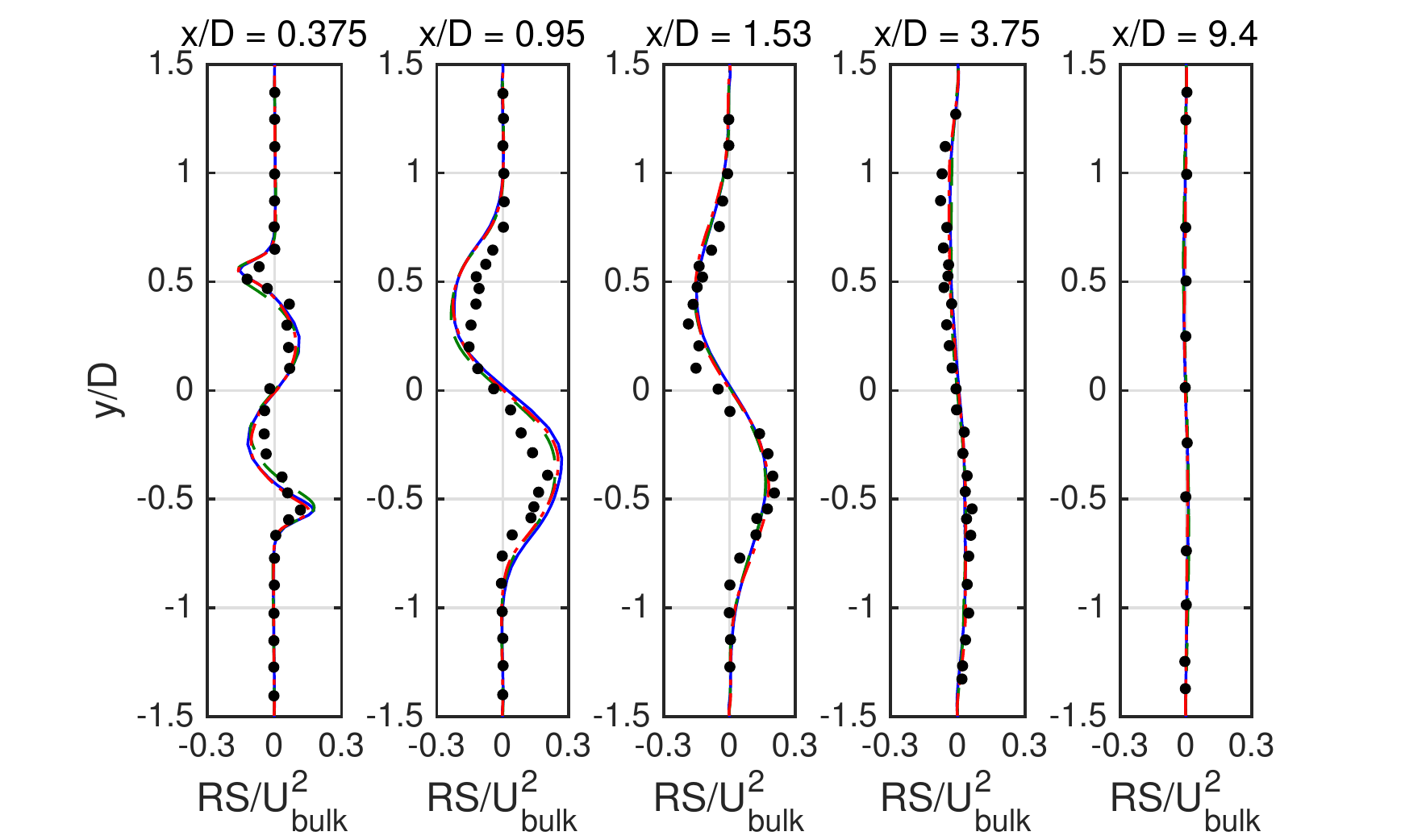}
    \caption{Normalized mean Reynolds stress profiles on three grids (solid line~\textendash\,~4~mm, dashed line~\textendash\,~2~mm, dash-dotted line~\textendash\,~1~mm) in comparison with measurements (dots), at several axial locations for the non-reacting case from the FV solver. \label{fig:reynolds}}
\end{figure}

\begin{figure}
    \centering
    \subfigure[Axial velocity]{
        \includegraphics[trim={0 0 0 0},width=0.48\textwidth,clip]{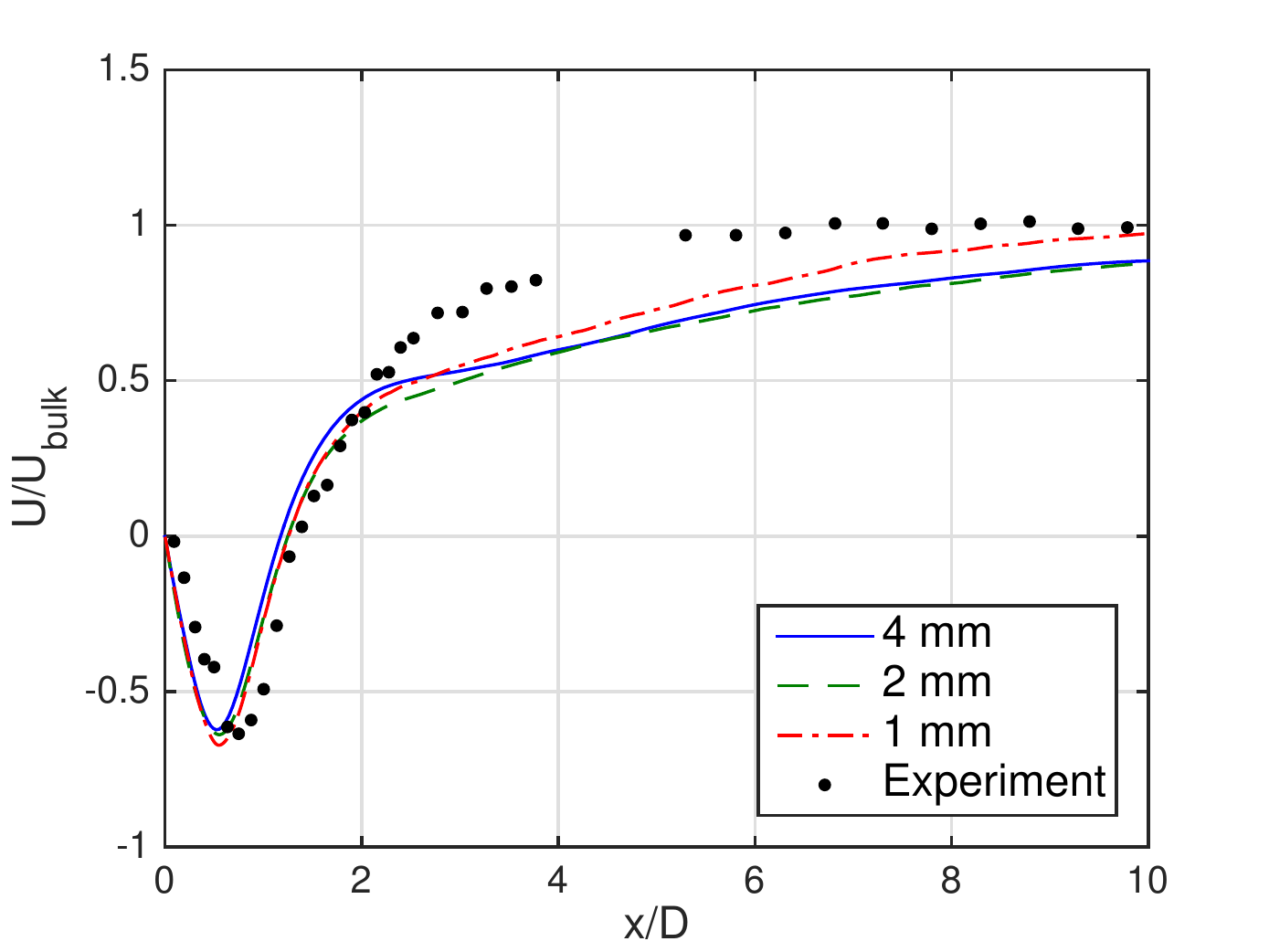}
    }
    \subfigure[Anisotropy]{
        \includegraphics[trim={0 0 0 0},width=0.48\textwidth,clip]{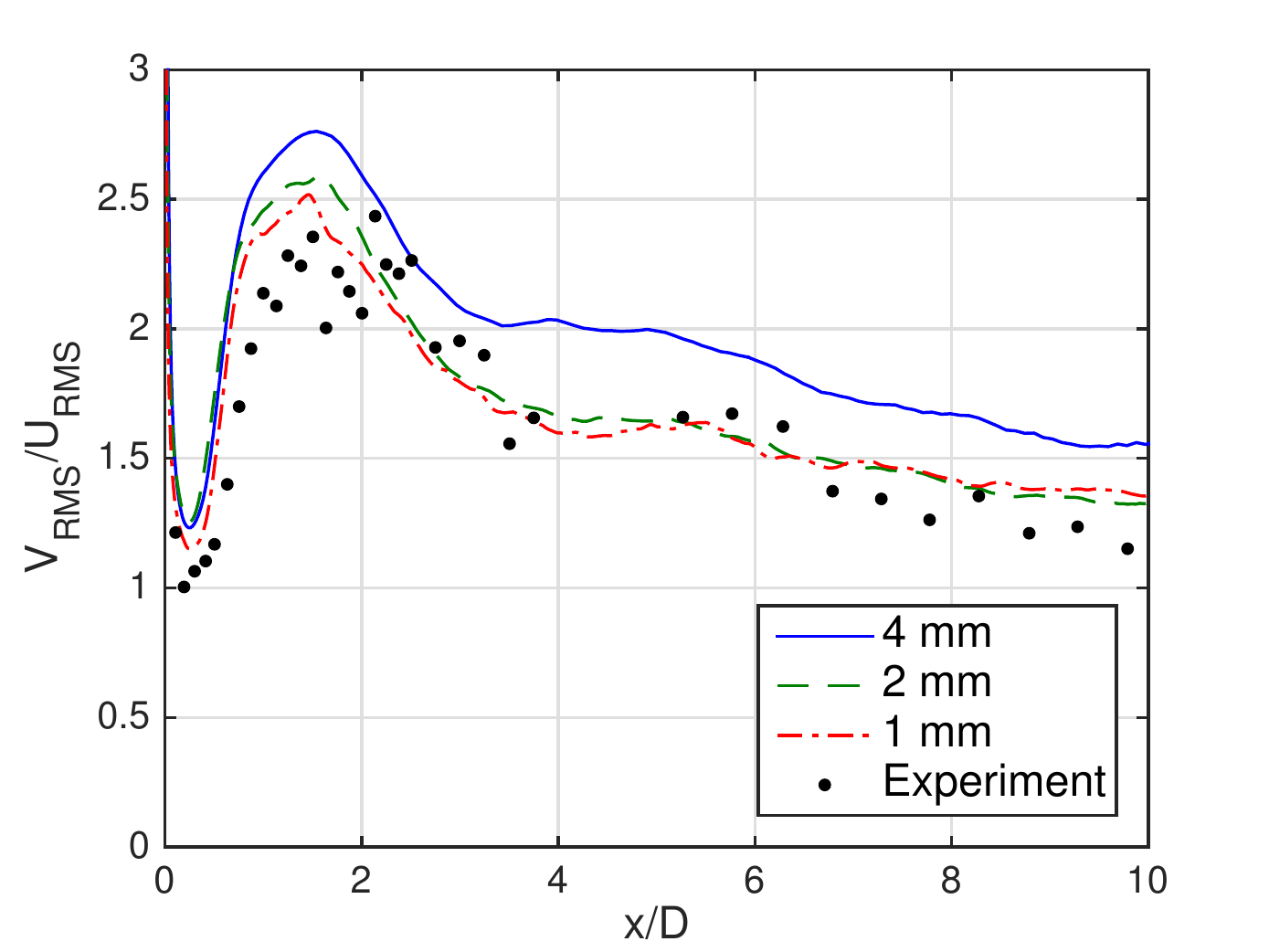}
    }
    \subfigure[Fluctuation]{
        \includegraphics[trim={0 0 0 0},width=0.48\textwidth,clip]{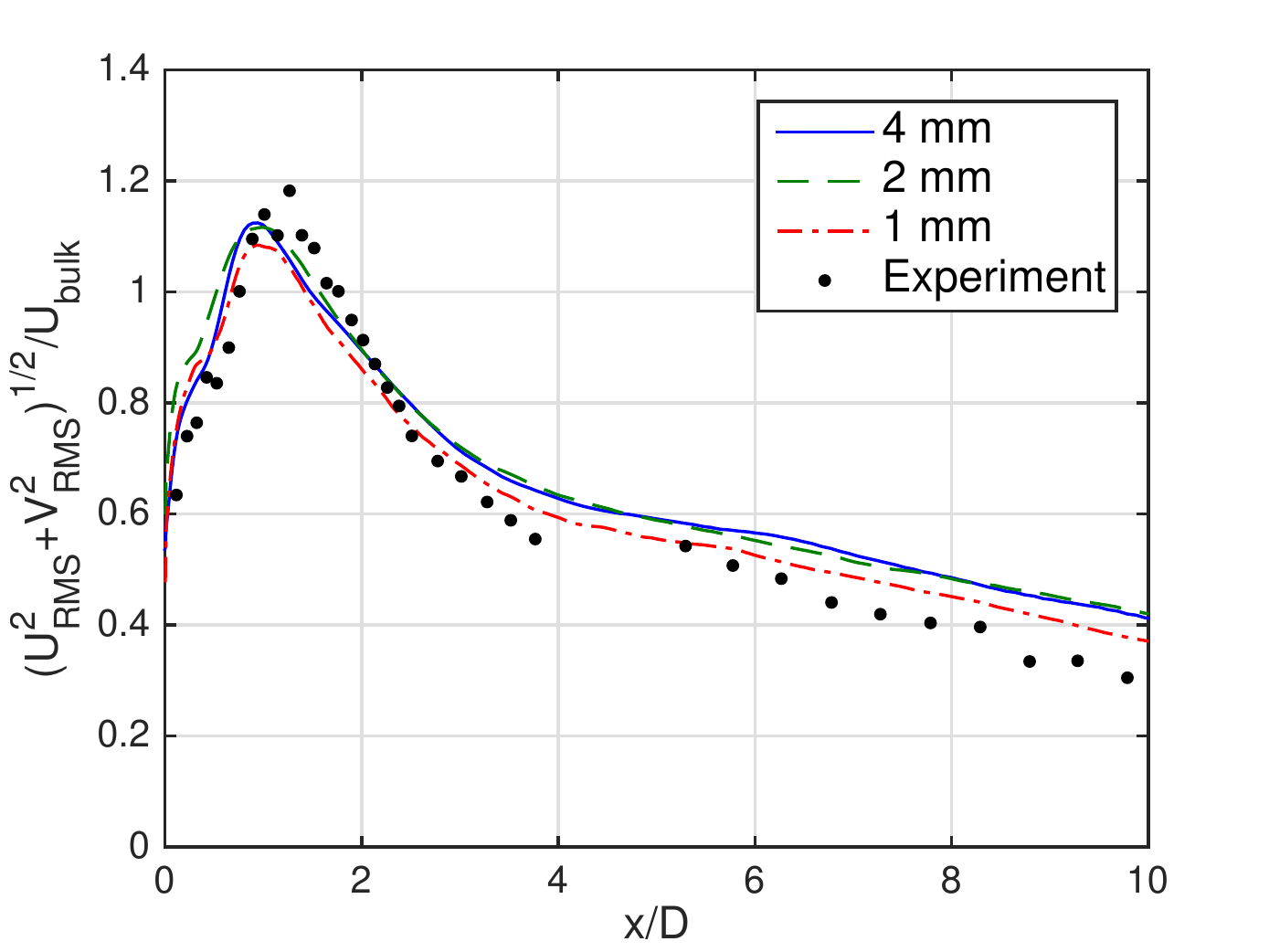}
    }
    \caption{Centerline profiles for (a) mean axial velocity, (b) anisotropy, and (c) fluctuation level on three grids (solid line~\textendash\,~4~mm, dashed line~\textendash\,~2~mm, dash-dotted line~\textendash\,~1~mm) in comparison with measurements (dots) for the non-reacting case from the FV solver. \label{fig:axialprofile}}
\end{figure}

\Crefrange{fig:velocityu}{fig:reynolds} show the results of the velocity statistical profiles for the three different grids used. The simulations are averaged for about 300~ms, which corresponds to about six flow-through times over the entire computational domain. Five different axial locations are considered for comparison with the experimental measurements. Mean and root-mean-square (RMS) axial and transverse velocity profiles are shown in \cref{fig:velocityu,fig:velocityv}, and the Reynolds stress profiles are shown in \cref{fig:reynolds}. It can be seen that grid convergence is achieved already with the 2~mm grid for both the first and second moment statistics, and the simulation with the 1~mm grid confirms the convergence. Excellent agreement can be observed for all the quantities compared with the measurements, except for the transverse velocity at the axial location $x/D = 1.53$, where the experiments have larger magnitude.

To further compare the simulation results with the experimental data quantitatively, two more quantities are considered together with the axial velocity at the centerline of the domain downstream the flameholder. The anisotropy and the fluctuation level are defined as $U_\text{RMS} / V_\text{RMS}$ and $\sqrt{U_\text{RMS}^2 + V_\text{RMS}^2} / U_\text{bulk}$, respectively. \Cref{fig:axialprofile} shows the results from simulations using the three grids. It be seen that although the results from three grid resolutions exhibits little difference for the axial velocity and fluctuation level. The 4~mm results for the anisotropy shows behavior quite different to that obtained for the other two grids. This shows that the 2~mm grid resolution is sufficient for the prediction of the instabilities and turbulence for the non-reacting case.

It can be seen from \cref{fig:axialprofile} that, although the anisotropy and fluctuation levels show excellent agreement with the experimental measurements, the axial velocity shows 20\% to 30\% underprediction for axial locations after $x/D = 2$. This discrepancy from the experimental measurements may be due to the different boundary conditions employed in the current simulation to that in the experiments, e.g. the flow is assumed to be laminar at the inlet. Similar underprediction of the axial velocity can be found in numerical simulations using other solvers with similar computational domain and boundary conditions~\cite{cocks2013reacting}. Overall, the non-reacting simulations from the FV solver show very good agreement with the experiments for the available data considered. 

\subsubsection{Reacting simulations}
The reacting flow simulations with the FV discretization is carried out using the 4mm-resolution, 2mm-resolution, and 1mm-resolution grids. Both simulations use the same combustion model as discussed in Sec.~\ref{SEC_NUM}. 

\begin{figure}
    \centering
    \includegraphics[width=0.8\textwidth]{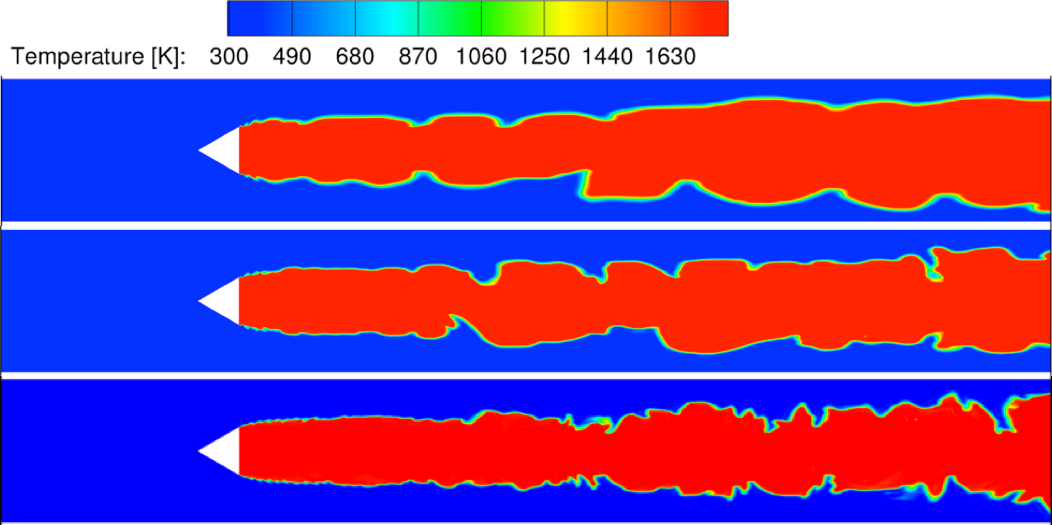}
    \caption{ Instantaneous temperature contours of 4mm (top), 2mm (middle) and 1mm (bottom) cases from the FV solver. \label{fig:inst_temp_R}}
\end{figure}

\begin{figure}
    \centering
    \includegraphics[trim={30 0 40 0},width=0.48\textwidth,clip]{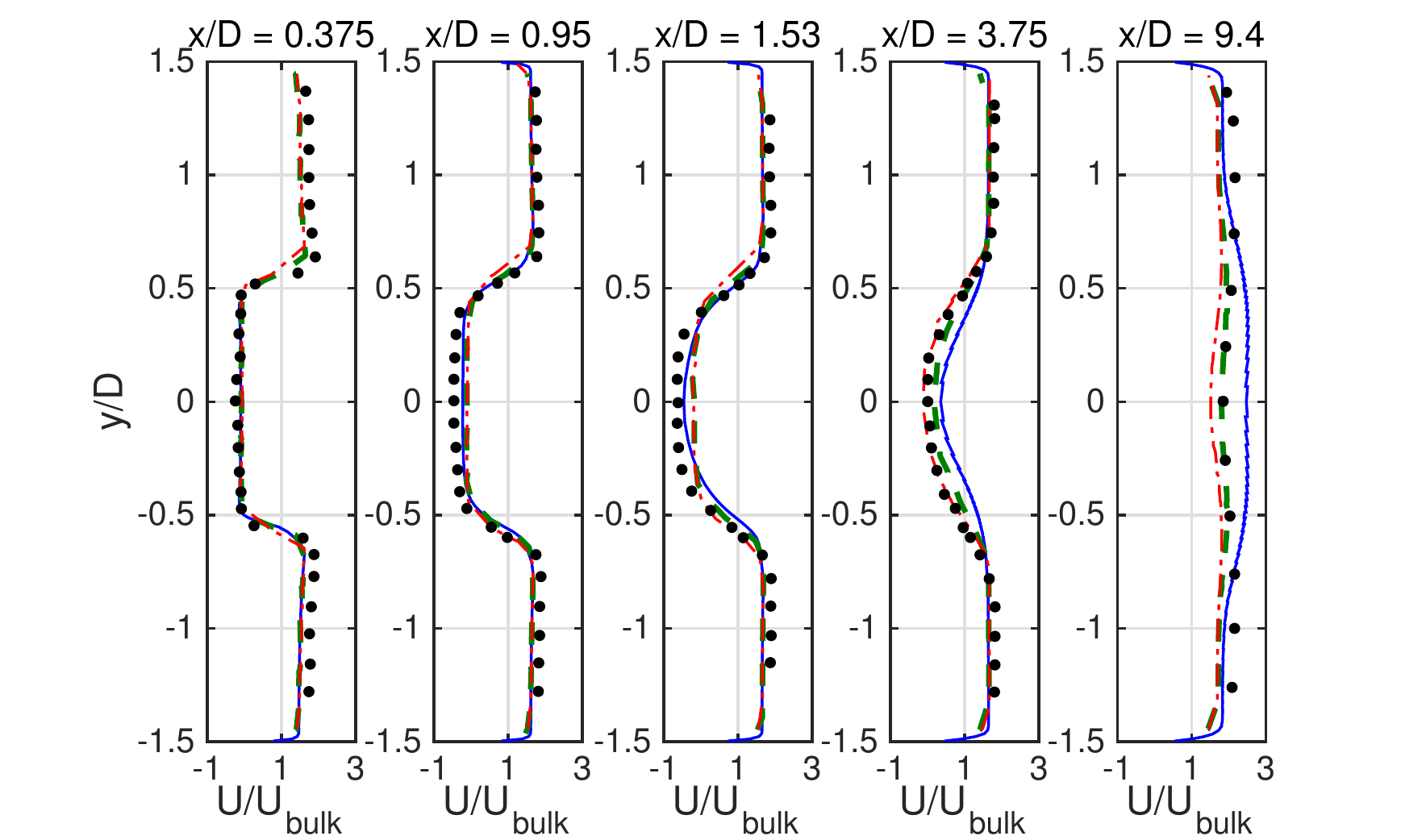}
    \includegraphics[trim={30 0 40 0},width=0.48\textwidth,clip]{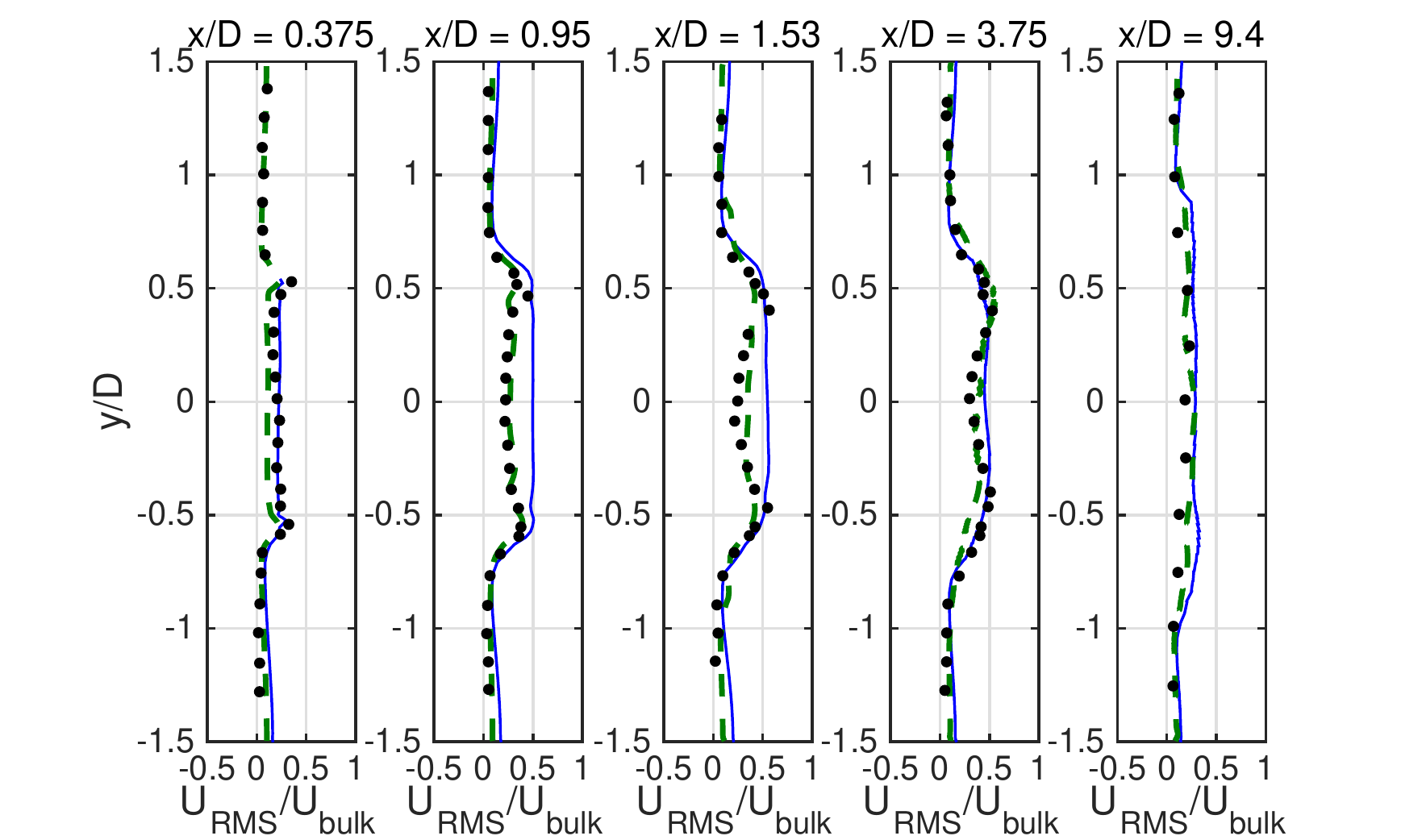}
    \caption{Normalized mean (left) and RMS (right) axial velocity profiles on three grids (solid line~\textendash\,~4~mm, dashed line~\textendash\,~2~mm in comparison with measurements (dots), dash-dotted line~\textendash\,~1~mm), at several axial locations for the non-reacting case from the FV solver. \label{fig:velocityu_R}}
\end{figure}

\begin{figure}
    \centering
    \includegraphics[trim={30 0 40 0},width=0.48\textwidth,clip]{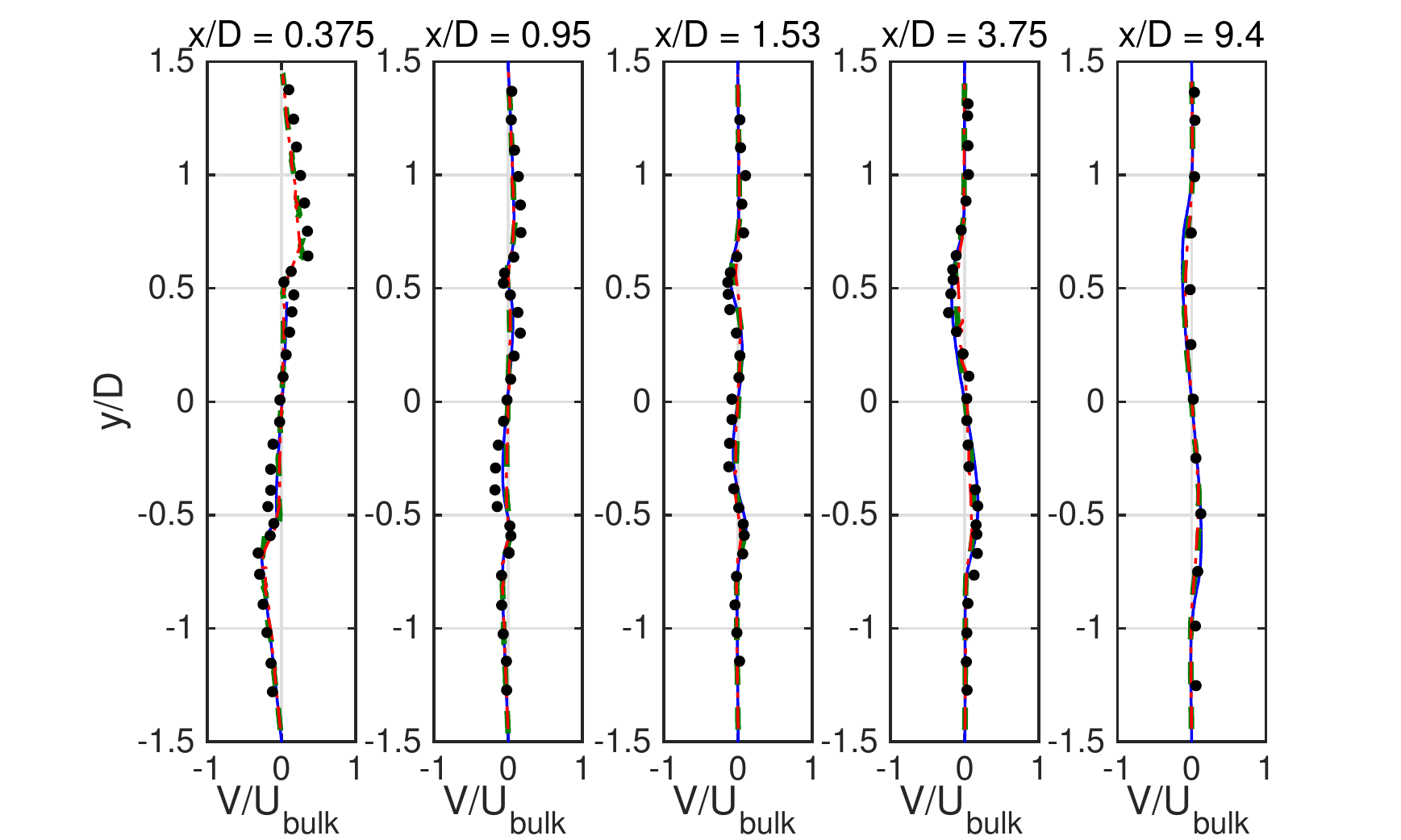}
    \includegraphics[trim={30 0 40 0},width=0.48\textwidth,clip]{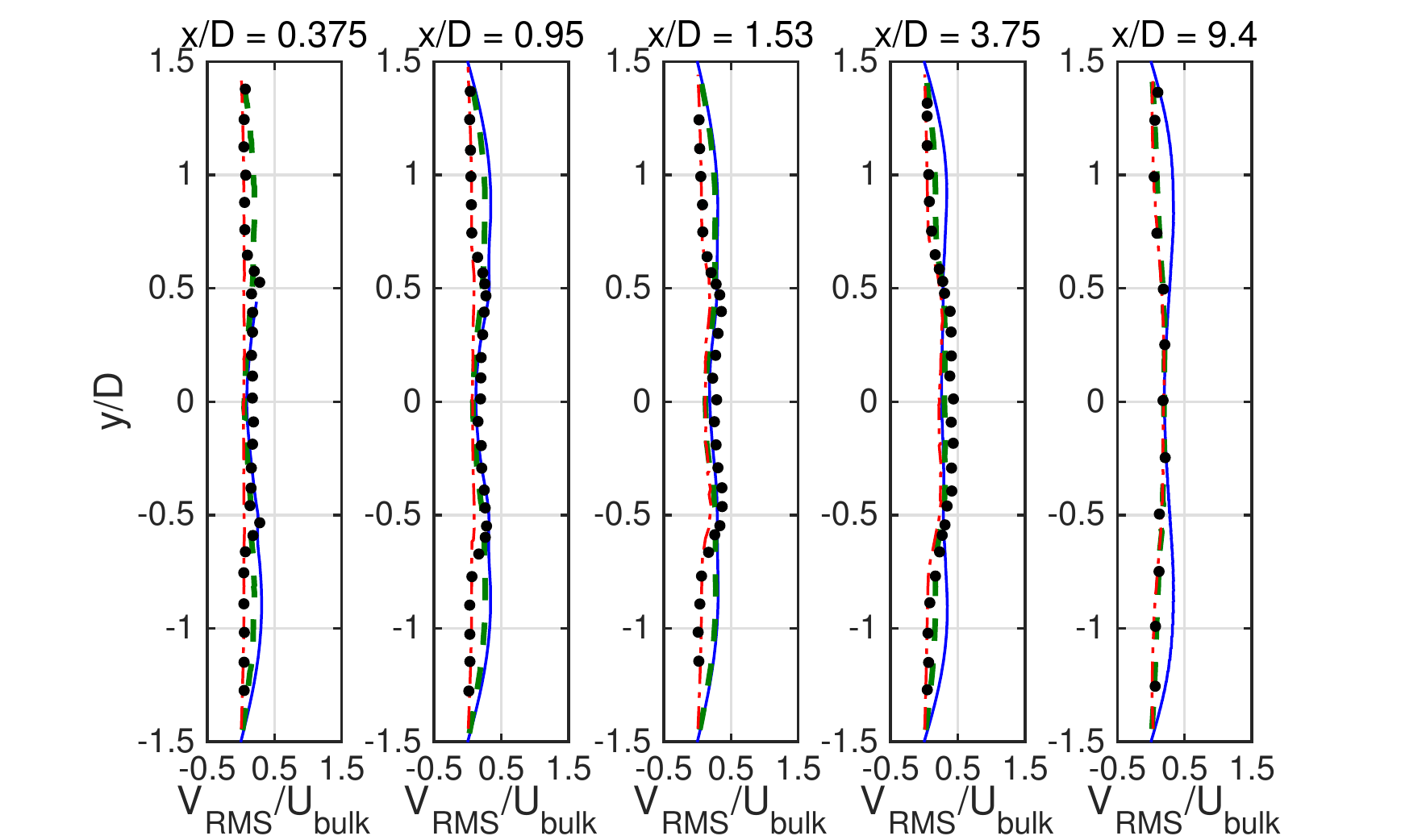}
    \caption{Normalized mean (left) and RMS (right) transverse velocity profiles on three grids (solid line~\textendash\,~4~mm, dashed line~\textendash\,~2~mm, dash-dotted line~\textendash\,~1~mm) in comparison with measurements (dots), at several axial locations for the non-reacting case from the FV solver. \label{fig:velocityv_R}}
\end{figure}
\begin{figure}
    \centering
    \includegraphics[trim={30 0 40 0},width=0.48\textwidth,clip]{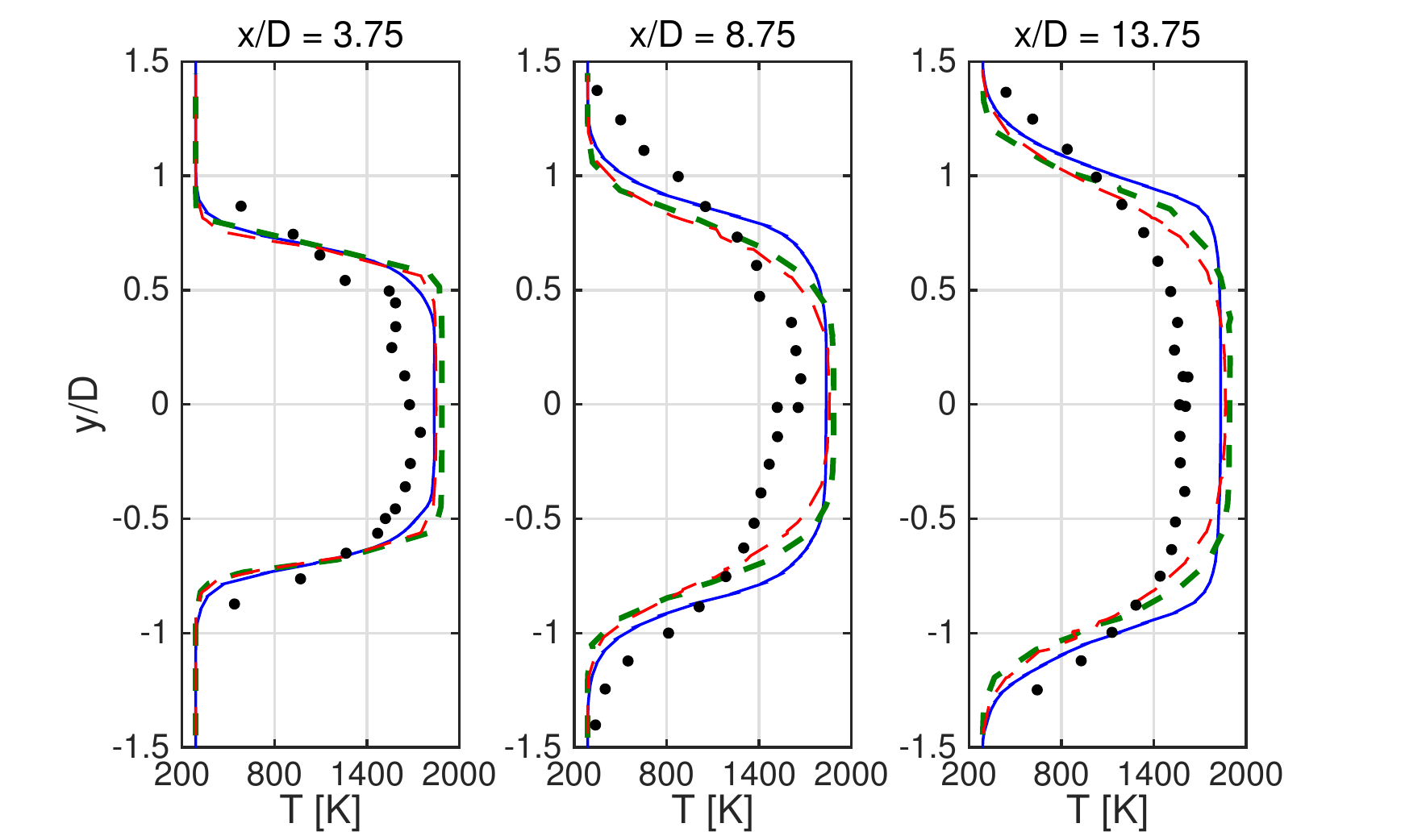}
    \caption{Mean temeparture profiles on three grids (solid line~\textendash\,~4~mm, dashed line~\textendash\,~2~mm, dash-dotted line~\textendash\,~1~mm) in comparison with measurements (dots), at several axial locations for the non-reacting case from the FV solver. \label{fig:temp_R}}
\end{figure}

\begin{figure}[!t!]
    \centering
    \subfigure[Axial velocity]{
        \includegraphics[trim={0 0 0 0},width=0.48\textwidth,clip]{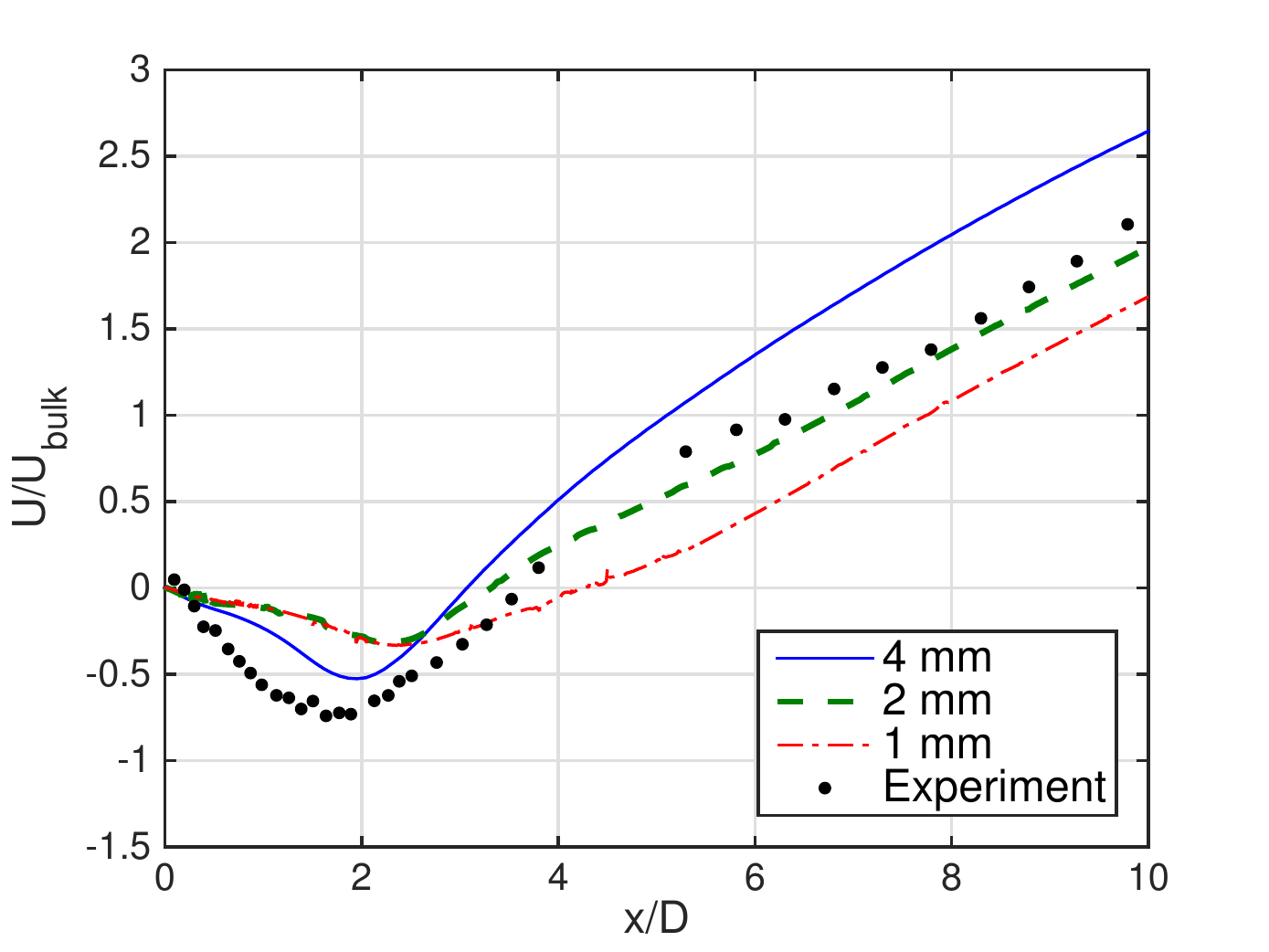}
    }
    \subfigure[Anisotropy]{
        \includegraphics[trim={0 0 0 0},width=0.48\textwidth,clip]{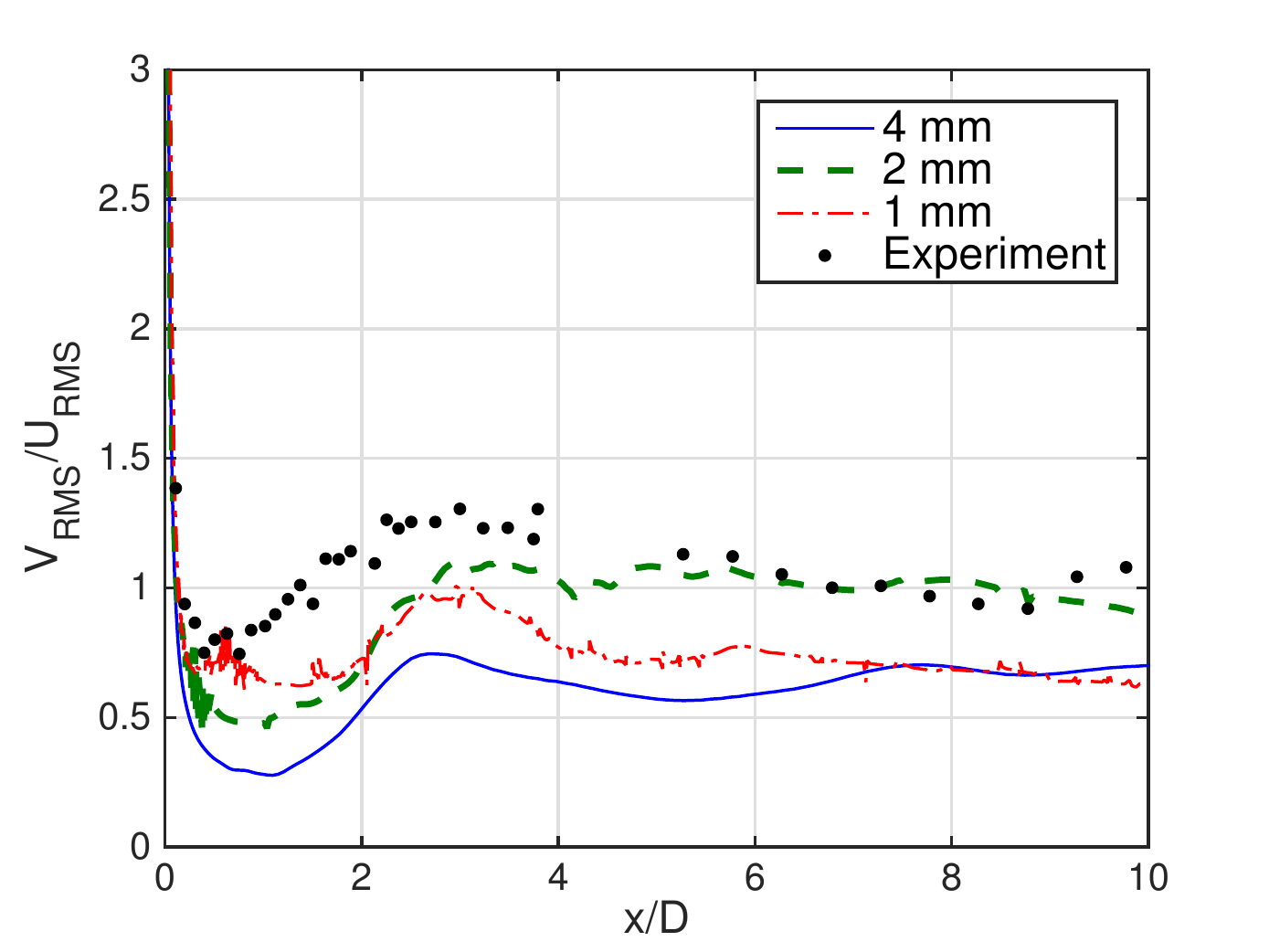}
    }
    \subfigure[Fluctuation]{
        \includegraphics[trim={0 0 0 0},width=0.48\textwidth,clip]{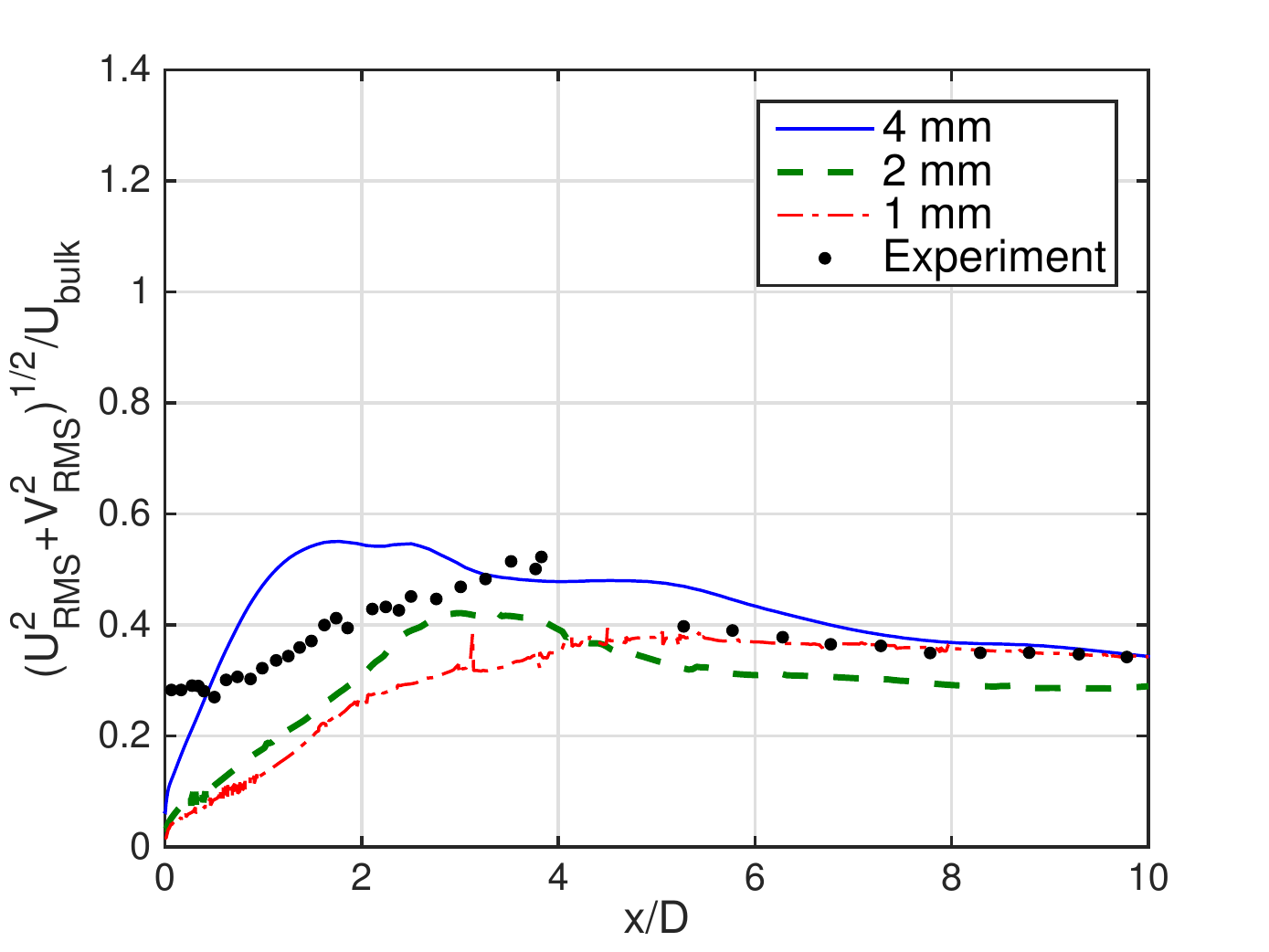}
    }
    \caption{Centerline profiles for (a) mean axial velocity, (b) anisotropy, and (c) fluctuation level on three grids (solid line~\textendash\,~4~mm, dashed line~\textendash\,~2~mm, dash-dotted line~\textendash\,~1~mm) in comparison with measurements (dots) for the non-reacting case from the FV solver. \label{fig:axialprofile_R}}
\end{figure}

The elevated temperature in the reacting case as well as the density ratio significantly alter the flow field characteristics. The flow is less turbulent due to the higher viscosity and the large density ratio suppresses sinuous instability mode, which is characteristic for non-reacting flow in the wake of a bluff-body. The laminar flame speed of the mixture with $\phi = 0.65$ is estimated to be $0.2$ m/s and the corresponding flame thickness is $0.6$ mm. Therefore, none of the three FV meshes calculations can resolve the flame and the impact of the flame thickening model on the simulation result is not negligible.

The instantaneous temperature fields of the 4mm and 2mm simulations are shown in Fig.~\ref{fig:inst_temp_R}. Vortex shedding is not present in either simulation. The instability of the shear layer at the near field of the wake is very symmetric. Intermittent sinuous modes can be observed at locations that are further downstream. A stronger effect of flame wrinkling can be observed in the 2mm case, which is not surprising given the higher resolution.

As part of the validation, the mean and RMS axial and transverse velocities across the flame are presented in Fig.~\ref{fig:velocityu_R} and Fig.~\ref{fig:velocityv_R}. Reasonable agreement is obtained by both cases in terms of the mean velocity profiles. The fluctuations are better captured by the higher-resolution case. The achievement of grid convergence cannot be observed from the two cases presented.

The mean axial velocity, anisotropy, and fluctuation along the centerline are shown in Fig~\ref{fig:axialprofile_R}. The cases with finer mesh shows the improvement over all three quantities, but the overall agreement with the experimental data is worse in comparison to the non-reacting case. Grid convergence has not been reached for the series of reacting calculations. The 2~mm case shows overall best agreement with respect to the experimental data among the three cases.

\subsection{Lyapunov exponent}
\begin{figure}
    \centering
    \includegraphics[width=0.5\textwidth]{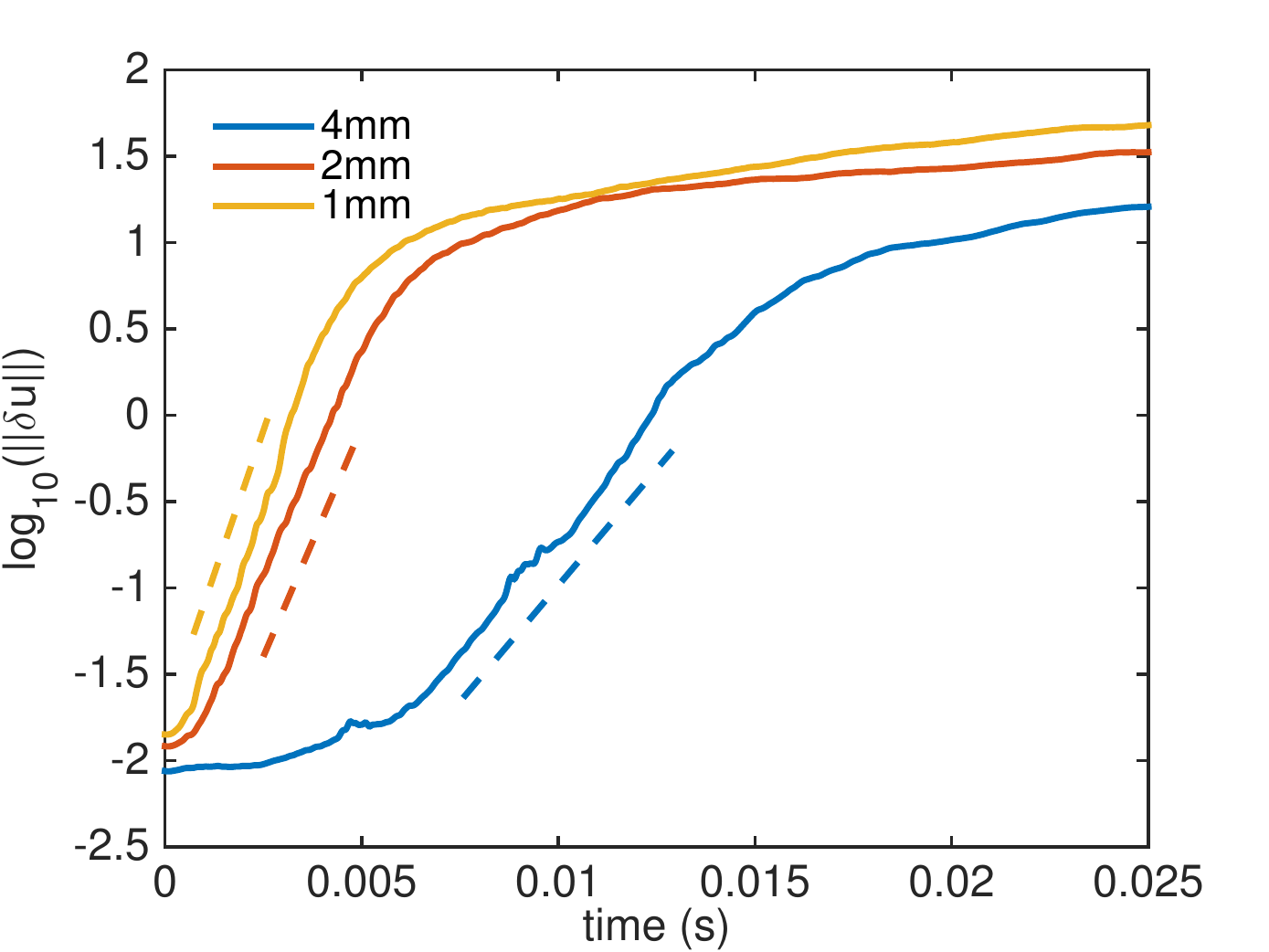}
    \caption{Separation and Lyapunov exponents as a function of grid resolution for the non-reacting simulations from the FV solver. The Lyapunov exponents is visualized in dashed lines, with $\lambda_{4mm} = 285$ s$^{-1}$, $\lambda_{2mm} = 517$ s$^{-1}$, $\lambda_{1mm} = 589$ s$^{-1}$. \label{fig:lyap_cold}}
\end{figure}

\begin{figure}
    \centering
    \subfigure[Axial velocity]{
        \includegraphics[trim={0 0 0 0},width=0.48\textwidth,clip]{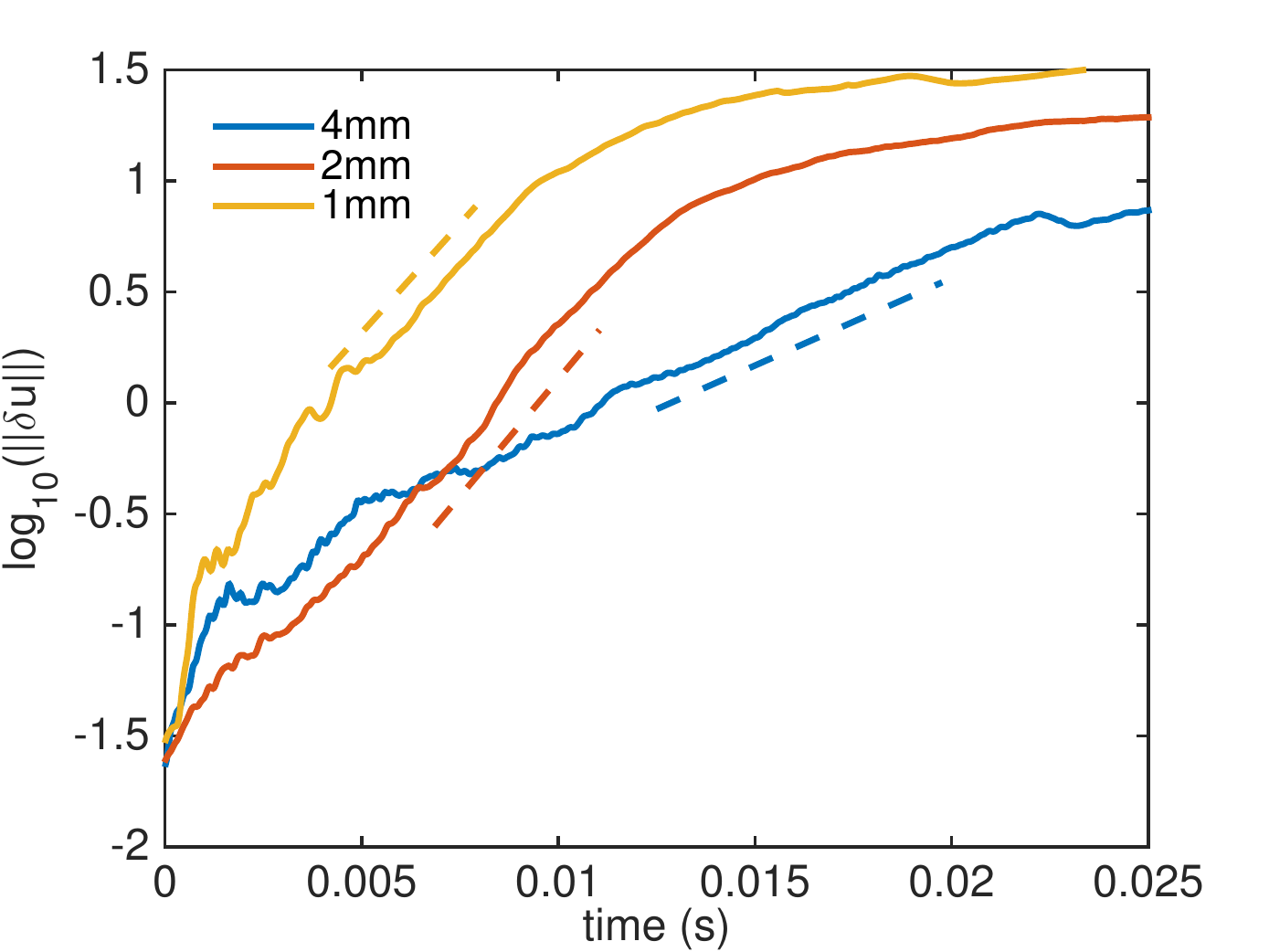}
    }
    \subfigure[Temperature]{
        \includegraphics[trim={0 0 0 0},width=0.48\textwidth,clip]{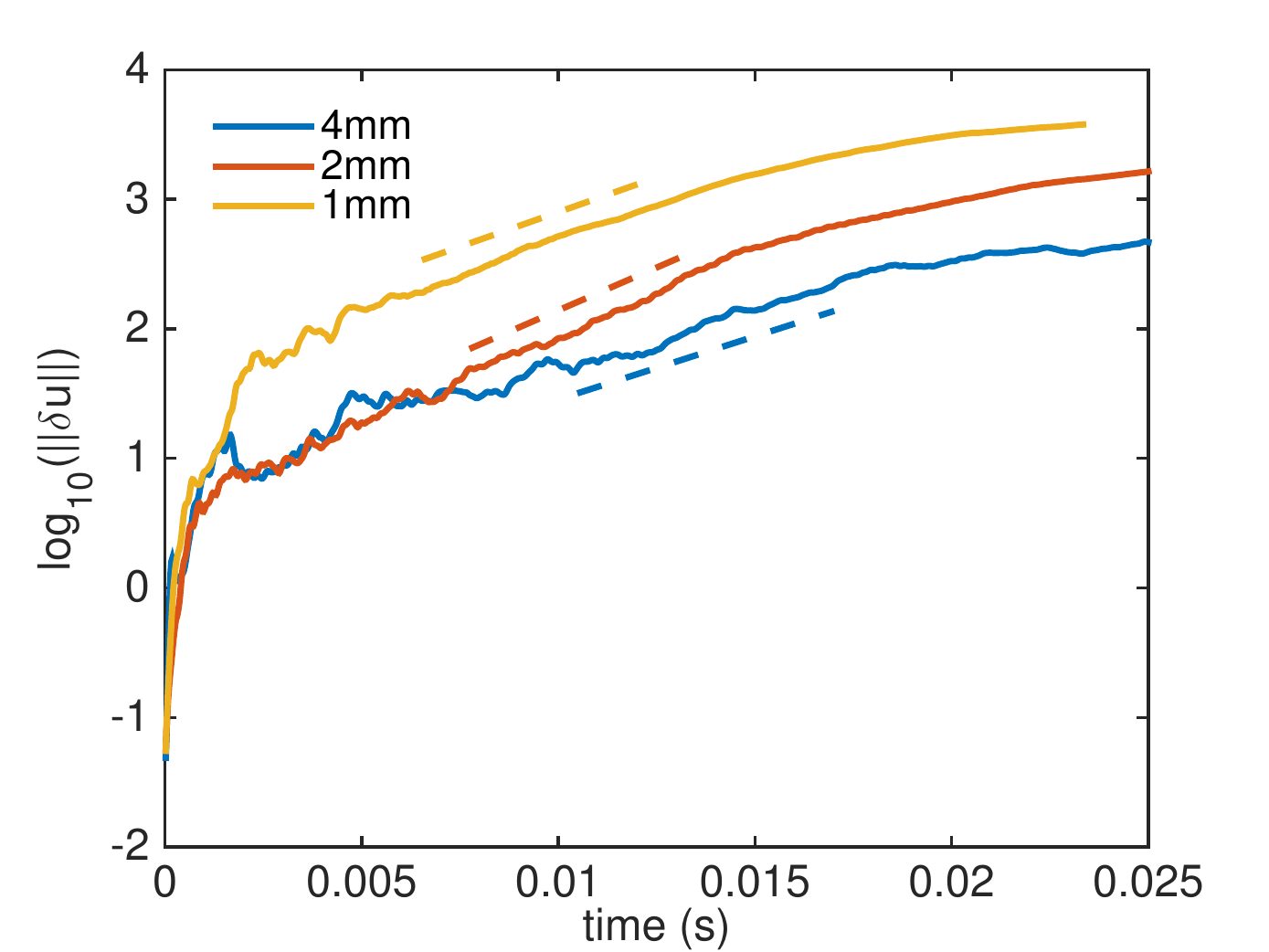}
    }
    \caption{Separation and Lyapunov exponents as a function of grid resolution for the reacting simulations with (a) axial velocity and (b) temperature. The Lyapunov exponents are $\lambda_{4mm} = 95$ s$^{-1}$, $\lambda_{2mm} = 206$ s$^{-1}$, $\lambda_{1mm} = 127$ s$^{-1}$ for the axial velocity, and $\lambda_{4mm} = 85$ s$^{-1}$, $\lambda_{2mm} = 119$ s$^{-1}$, $\lambda_{2mm} = 111$ s$^{-1}$ for temperature. \label{fig:lyap_hot}}
\end{figure}

The Lyapunov exponents are calculated for the reacting and non-reacting cases with different mesh resolutions. The perturbation is applied to the axial velocity with the initial relative magnitude of $\epsilon = 1.0 \times 10^{-4}$. The resulting growth of the separation between the perturbed and unperturbed simulations are shown in Fig.~\ref{fig:lyap_cold} and~\ref{fig:lyap_hot}. For the non-reacting cases, the separation and Lyapunov exponents of the axial velocity are calculated for the cases using $4$-mm, $2$-mm, and $1$-mm mesh resolutions. The Lyapunov exponents are calculated to be $\lambda_{4mm} = 285$ s$^{-1}$, $\lambda_{2mm} = 517$ s$^{-1}$, $\lambda_{1mm} = 589$ s$^{-1}$. Expectedly, the growth rate increases as the mesh becomes finer and more chaotic dynamics are resolved. Moreover, the difference of the slope decreases for more resolved calculations. The convergence in the Lyapunov exponents agrees well with the observed convergence of the corresponding statistics.

For the reacting cases, Lyapunov exponents for the axial velocity as well as the temperature are obtained for the $4$-mm, $2$-mm, and $1$-mm mesh resolutions. As shown in Fig.~\ref{fig:lyap_hot}, for the reacting cases, slower growth of the separation results in the lower values of Lyapunov exponent compared with the non-reacting cases. The exponents for the axial velocity are calculated to be $\lambda_{4mm} = 95$ s$^{-1}$, $\lambda_{2mm} = 206$ s$^{-1}$, and $\lambda_{1mm} = 127$ for the $4$-mm, $2$-mm, and $1$-mm mesh resolutions respectively. For temperature, the growth rates are calculated to be $\lambda_{4mm} = 85$ s$^{-1}$, $\lambda_{2mm} = 119$ s$^{-1}$, $\lambda_{2mm} = 111$ s$^{-1}$. The change of the Lyapunov exponent with refined mesh is more significant for the axial velocity, indicating a stronger mesh-dependency for the velocity compared with the temperature. Grid convergence has not be reached for the reacting simulations as revealed by the difference in the trajectories of the separation.

\subsection{Wasserstein metric}
\begin{figure}
    \centering
    \includegraphics[width=0.6\textwidth]{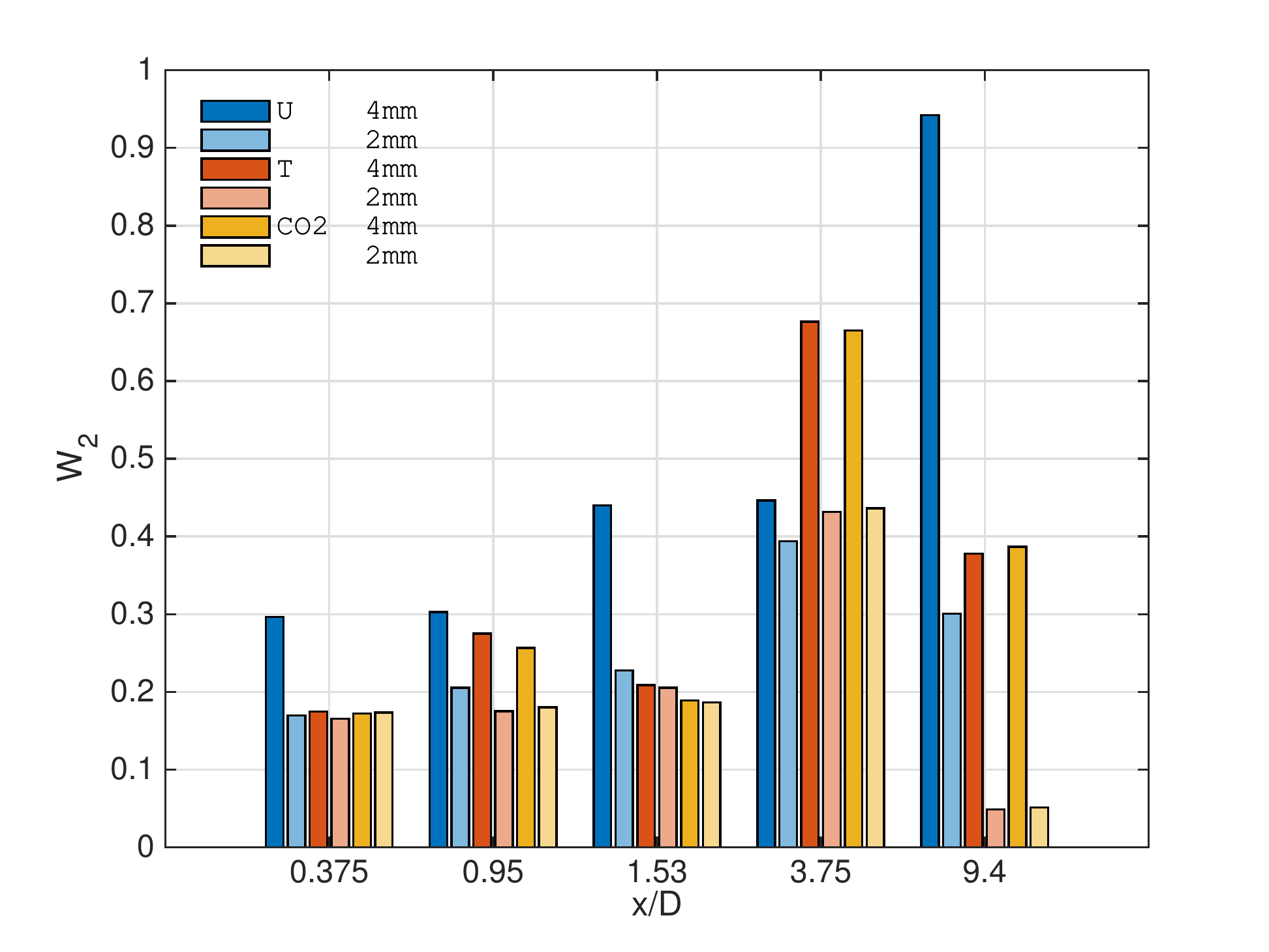}
    \caption{The Wasserstein metric ($W_2$) for the axial velocity, the temperature, and the mass fraction of $\ce{CO_2}$ obtained from the $4$-mm and $2$-mm cases, with respect to their correspondence of the $1$-mm case, at four axial locations. \label{fig:W2}}
\end{figure}

For the reacting cases, the difference between simulations with various grid resolutions can be further quantified by the Wasserstein metric ($W_2$), using the axial velocity, the temperature, and the mass fraction of $\ce{CO_2}$ obtained from the $4$-mm and $2$-mm cases, are compared with those of the $1$-mm case. The values of $W_2$ at five axial locations are displayed in Fig.~\ref{fig:W2}. The metric is normalized with respect to the standard deviations obtained from the $1$-mm case for each quantity. Consistent decreases in the value of $W_2$ can be observed for all quantities of interest, indicating the reduction of difference between the simulations results as the resolution increases. The deviation increases along the axial direction due to the accumulation of the error. The difference between $W_2$ is most significant for the axial velocity and the species mass fractions. The strong mesh-dependency for the axial velocity in comparison with the temperature agrees well with the observation made from the analysis of the Lyapunov exponents.

\section{Conclusions} \label{SEC_CONC}
In this work, methods for the evaluation of LES-quality and LES-accuracy are presented, which include the Lyapunov exponent for the analysis of short-time predictability of LES-calculation and the Wasserstein metric for the quantitative assessment of simulation results. Both methods are derived and evaluated in application to the Volvo test case. Both the non-reacting and reacting cases are calculated. For the non-reacting cases, good agreement with the experimental data is achieved by solvers at high numerical resolution. The reacting cases are more challenging due to the small length scale of the flame and the suppression of sinuous mode of absolute instability by the density ratio. The analysis of the turbulent simulation data using the concept of the Lyapunov exponent and the Wasserstein metric provides a more quantitative approach to assess the mesh dependency of the simulation results. The convergence of the Lyapunov exponent is shown to be a more sensitive and stronger indication of mesh-independence. Though grid convergence for the reacting cases cannot be reached with the chosen resolutions, the Lyapunov exponents and the Wasserstein metric are shown to be capable of identifying quantity-specific sensitivities with respect to the numerical resolution, while requiring significantly less computational resources than acquiring profiles of conventional turbulent statistics.

\section*{Acknowledgments}
This work was funded by NASA Transformational Tools and Technologies Project with Award No. NNX15AV04A.

\bibliography{bibliography,bibliography_DG,lyapunovExp,wasserstein,emd,stats,combustion}
\bibliographystyle{aiaa}

\end{document}